\documentstyle[amssymb,psfig]{l-aa}

\begin{document}

\thesaurus{06(02.08.01; 08.02.1; 08.14.1; 13.07.1)}

\title{Mass ejection in neutron star mergers}

\author{S. Rosswog \inst{1}
\and M. Liebend\"orfer \inst{1}
\and F.-K. Thielemann \inst{1}
\and M.B. Davies \inst{2}
\and W. Benz \inst{3}
\and T. Piran \inst{4}}

\institute{Departement f\"ur Physik und Astronomie, Universit\"at Basel, 
            Switzerland
	   \and Institute of Astronomy, University of Cambridge, UK
           \and Physikalisches Institut, Universit\"at  Bern, Switzerland
           \and Racah Institute for Physics, Hebrew University, Jerusalem, 
		Israel}

\offprints{S. Rosswog, \\ email: rosswog@quasar.physik.unibas.ch, \\
                          Fax: **41 61 267 3784}

\date{Received ; accepted }

\maketitle

\begin{abstract}
We present the results of 3D Newtonian SPH simulations of the merger of a neutron star binary. The microscopic properties of matter are described by the physical equation of state of Lattimer and Swesty (LS-EOS). To check the model dependence of the results we vary the resolution ($\sim 21000$ and $\sim 50000$ particles), the equation of state (stiff and soft polytropes), the artificial viscosity scheme, the stellar masses, we include neutrinos (free-streaming limit), switch off the gravitational backreaction force, and vary the initial stellar spins. In addition we test the influence of the initial configuration, i.e. spherical stars versus corotating equilibrium configurations. The final matter distribution consists of a rapidly spinning central object with 2.5 to 3.1 ${\rm M}_{\odot}$ of baryonic mass that probably collapses to a black hole, a thick disk of 0.1 to 0.3 ${\rm M}_{\odot}$ and an extended low density region. In the case of corotation this low density material forms spiral arms that expand explosively due to an increase of the adiabatic exponent and the release of nuclear binding energy in the case of the LS-EOS, but remain narrow and well defined for the stiff polytropic equation of state. The main and new result is that for the realistic LS-EOS, depending on the initial spin, between $4\cdot10^{-3}$ and $4\cdot10^{-2}$ ${\rm M}_{\odot}$ of material become unbound. If, as suggested, large parts of this matter consist of r-process nuclei, neutron star mergers could account for the whole observed r-process material in the Galaxy.
\end{abstract}


\section{Introduction}
The question of the origin of Gamma Ray Bursts (GRBs) -whether they are galactic or cosmological- has been controversial during the three decades since the announcement of their detection (Klebesadel et al. 1973). In spite of the fact that  the Burst and Transient Source Experiment (BATSE) detected bursts at a rate of around one per day, no clear indication of a distance scale could be found.
It was only recently that  observations of the afterglow of Gamma Ray Bursts in the optical (van Paradijs et al. 1997; Sahu  et al. 1997; Galama  et al. 1997; Djorgovski  et al. 1997), X-ray (Costa  et al. 1997) and radio (Frail et al. 1997) part of the spectrum seem to have settled their cosmological origin. The identification of Fe and Mg absorption lines and the determination of the cosmological redshift ($0.835 \le z \le 2.3$; Metzger et al.  1997) seems to have definitely ruled out galactic sources of GRBs. The most popular model for the central engine to power cosmological GRBs is the merger of two neutron stars (or a neutron star and a black hole; see e.g. Paczy$\acute{{\rm n}}$ski 1986, 1991, 1992; Eichler et al. 1989; Rees $\&$ Mes$\acute{{\rm c}}$z$\acute{{\rm a}}$ros 1992; Narayan et al. 1992).\\
This scenario is also attractive for other reasons. By now five such neutron star binary systems are known (Thorsett 1996) and no mechanism is known that could prevent the inspiral and  final coalescence. During the last $\sim$ 15 minutes of its inspiral the binary system will emit gravitational waves with a frequency ranging from $\sim$ 10 Hz to $\sim$  1000 Hz. The inspiral is thus one of the prime candidates to be detected by earth bound gravitational wave detector facilities such as  LIGO (Abramovici et al. 1992), VIRGO (Bradaschia et al. 1990)  or GEO600 (Luck 1997).
Another reason to study this scenario -and this will be our main focus here- is the  question of nucleosynthesis, namely the production site of r-process nuclei. Despite considerable efforts, it has up to now not been possible to identify  the corresponding astrophysical event unambiguously. Most recent nucleosynthesis studies (see Freiburghaus et al. 1997, 1998; Meyer \& Brown 1997) raise questions concerning the ability of high entropy neutrino wind scenarios  in type II supernovae to produce r-process nuclei for $A< 110$ in correct amounts. In addition, it remains an open question whether the  entropies required for the nuclei with $A>110$ can actually be attained in type II supernova events. Thus, an alternative or supplementary low entropy, low $Y_{e}$
r-process environment seems to be needed (decompression of neutron star material).\\
The decompression of neutron star matter as a possible source of r-process nuclei was first discussed by Lattimer $\&$ Schramm (1974, 1976) in their study of the tidal disruption of a neutron star by a black hole. The coalescence of two neutron stars has in this context been examined by Symbalisty $\&$ Schramm 
(1982) and Eichler et al. (1989). Detailed studies of the nuclear decompression process have been performed by Meyer (1989).\\
Hydrodynamic simulations of coalescing  neutron star binaries have been performed by various groups. The first calculations were done by Nakamura and Oohara (see Shibata et al. (1993) and references therein) using polytropic equations of state and focusing mainly on the emitted gravitational waves.\\
Similar calculations, but using SPH, have been performed by Davies et al. (1994) who discussed a variety of physical effects related with the merging event. 
  Zhuge et al. (1994, 1996) focussed in their work  on the gravitational wave energy spectra $dE/df$.\\
In a long series of papers Lai, Rasio and Shapiro examined close binary systems. They developed an approximate analytical energy variational method and applied it to analyze the stability properties of binary systems and rotating stars (Lai et al. 1993a, 1993b, 1994a, 1994b, 1994c, Lai 1994). Two of them (Rasio and Shapiro) performed complementary SPH-simulations, where apart from stability questions the emission of gravitational waves was investigated (Rasio $\&$ Shapiro 1992, 1994, 1995).\\
Ruffert et al. (1996, 1997) performed PPM-simulations of neutron star mergers, using the Lattimer/Swesty EOS and accounting for neutrino emission by means of an elaborate leakage scheme. They discussed gravitational wave and neutrino emission  and made an attempt to address the question of mass ejection by looking at the material they lost from their grid.\\
Rosswog et al. (1998a,b) discussed preliminary results of their SPH calculations concerning mass ejection and its implications for nucleosynthesis.\\
Fully relativistic hydrodynamical calculations are beginning to yield results (see Wilson $\&$ Mathews 1995, Wilson et al. 1996, Mathews $\&$ Wilson 1997 , Baumgarte et al. 1997). However, these  are still controversial (see e.g. Lai 1996, Thorne 1997, Lombardi $\&$ Rasio 1997, Wiseman 1997, Shibata et al. 1998).\\
Our focus in this paper will be on the matter that becomes unbound in an equal mass neutron star binary merger. In Sect. 2 we describe the numerical method, we  discuss our chosen binary system parameters and the appropriateness of the approximations of our model. The results are discussed in Sect. 3. In 3.1 we describe the morphology of the mergers, in 3.2 the mass distribution is discussed, 3.3 deals with the thermodynamic properties  of the merged configuration, in 3.4 we discuss our neutrino treatment and 3.5 deals with mass ejection and related questions of nucleosynthesis. The results are summarized in Sect. 4. Details of the initial SPH-particle setup, the 
equation of state and the neutrino emission are given in the appendices.

\section{Numerical Method and Physical Parameters of the System}
\subsection{The Numerical Method}
We perform hydrodynamic calculations using a 3D Newtonian SPH code based on the one used in Davies et al. (1994). The basic equations of our SPH formulation can be found in the review of Benz (1990), the gravitational forces are calculated using a hierarchical binary tree (Benz et al. 1990). Since the method is well-known and comprehensive reviews exist (see e.g. Hernquist $\&$ Katz (1989); Benz (1990); Monaghan (1992); M\"uller (1998)), we do not go into the details of SPH here and refer the reader to the literature.\\
The Lagrangian nature of SPH is perfectly suited to the study of this intrinsically three dimensional problem, since it is not subject to spatial restrictions imposed by a computational grid. Thus, no matter can be lost from the computational domain (as for example in Eulerian methods). In addition no CPU resources are wasted for uninteresting empty regions. These zones are in other methods sometimes filled by low density material that can possibly lead to artifacts. Especially, the (tiny) amounts of ejecta could be influenced since they have to struggle against surrounding material. SPH also allows one to follow individual particles and therefore to keep track of the history of chosen blobs of matter.\\
The details of the initial SPH-particle setup are given in Appendix A.

\subsection{The Physical Parameters of the Binary System}
In the following we want to discuss the physical parameters of the binary system together with the chosen approaches and their uncertainties that will be tested.

\subsubsection{Masses}
Present state of the art nuclear equations of state allow gravitational masses for maximally rotating neutron stars of $\sim 2.5 \; {\rm M}_{\odot}$ (see e.g. Weber $\&$ Glendenning (1992)). However, all observed neutron star masses are centered in a narrow band around 1.4 ${\rm M}_{\odot}$ (e.g. van den Heuvel (1994)), a mass that might be given rather by evolutionary constraints than by the stability limit of the EOS. The masses that are found in the five known neutron star binaries are very close to each other, the maximum deviation being $\sim$ 4 \% in the case of PSR 1913+16 (see Thorsett 1996).
Considering the observational facts we only investigate equal mass binary systems. We regard a baryonic mass of 1.6 ${\rm M}_{\odot}$, corresponding approximately to 1.4 ${\rm M}_{\odot}$ of gravitational mass, as most reasonable.
The amount of mass that is ejected  into the interstellar medium is directly related to the gravitational potential that has to be overcome by the neutron star debris. Hence we vary the neutron star masses, investigating also the case of $1.4 \; {\rm M}_{\odot}$ of baryonic matter per star. Starting with the initial separation $a_{0}= 45$ km, which is our standard value, the inspiral will take longer in this case than for the heavier stars. For an estimate one may apply the formula for the inspiral time of a point mass binary (see e.g. Misner et al. 1973) 

\begin{equation}
\tau= \frac{5 \; c^{5}}{256 \; G^{3}} \frac{a_{0}^{4}}{\mu  M^{2}}\label{inspiral}
\end{equation}
giving an inspiral time that is longer by a factor of approximately $(8/7)^{3}\approx 1.5$ for the lighter case.

\subsubsection{Gravitational Radiation Backreaction}
Since the backreaction force, resulting from the emission of gravitational waves, tends to circularize elliptic orbits (see Peters $\&$ Mathews (1963, 1964)), we start with  basically circular orbits.
Starting on the x-axis of our coordinate system, we give each SPH-particle a tangential velocity corresponding to a circular Kepler motion with a fixed value $\omega^{2}= G M a_{0}^{-3}$, where $M$ denotes the total  mass of the system, and add the radial velocity of a point mass binary in quadrupole approximation (see e.g. Shapiro $\&$ Teukolsky 1983)
\begin{equation}
\frac{dr}{dt}=-
\frac{64}{5} 
\frac{G^{3}}{c^{5}} \frac{M_{j}^{3}}{a^{3}_{0}}
\end{equation}
($r= a/2$, $M_{j}$ is the mass of each star).
As in Davies et al. (1994) we apply in addition to the hydrodynamic and gravitational accelerations the backreaction of a point mass binary to each SPH-particle:

\begin{eqnarray}
\ddot{x}_{{\rm gwb}}= \frac{\eta}{2 \; M_{j} \;(\vec{v} \vec{r})} (E x + \frac{J \; \dot{y}}{2})\label{gwback1}\\
\vspace{0.8 cm}
\ddot{y}_{{\rm gwb}}= \frac{\eta}{2 \; M_{j} \;(\vec{v} \vec{r})} (E y - \frac{J \; \dot{x}}{2})\label{gwback2}.
\end{eqnarray}
Here $\eta= \frac{64}{5} \frac{G^{3}}{c^{5}} \frac{M^{2}\; \mu}{a^{4}}$, $E$ is the energy, $J$ the angular momentum, $\mu$ the reduced mass  and $\vec{v} , \;  \vec{r}, \; x, \; y$,  $\dot{x}$ and $\dot{y}$ refer to the point mass binary (not to be confused with the  SPH-particle properties). Since this acceleration is applied to each  particle, the circulation of the fluid will be conserved. When the distances of the centers of mass of the stars from the origin are equal to one stellar radius (and thus the point mass approximation definitely breaks down) the backreaction force is switched off.\\
This treatment of the backreaction force is clearly a simplification of the relevant gravitational physics. To test the corresponding sensitivity of our results we will investigate also the case where this force is totally neglected. To push the system over the critical radius where it becomes dynamically unstable (for an estimate of this radius obtained by applying  an energy variational method to compressible, triaxial ellipsoids obeying a polytropic equation of state see Lai et al. (1994a) and references therein), we have to begin in this case with a smaller  initial separation ($42$ km).

\subsubsection{Tidal Deformation}
For the  standard case of our calculation we will start with spherical initial configurations. However, a neutron star binary system that has spiraled down to a center of mass distance of $a_{0}= 45$ km will show non-negligible tidal deformations. In a crude estimate one finds for the height of the tidal bulge in our equal mass system $h\approx (\frac{R}{a_{0}})^{3} \cdot R \approx 0.5 $ km, i.e. parts of the surfaces of the neutron stars will be raised to up to half a kilometer above their spherical radius $R$. Thus starting with initially spherical stars will lead to oscillations. To estimate the impact of the approximation of spherical stars we compare the results with the case where the initial configuration has been relaxed in the mutual gravitational field. \\
In order to obtain a hydrostatic equilibrium for the tidally locked, corotating
 binary system, we consider the coordinate system in which the neutron stars
 are at rest.
A particle $i$ of mass $m_{i}$ is accelerated not only by the forces in the inertial frame $\vec{F}_{i}$, but also by non-inertial forces (see e.g. Landau 1976), so that the  total force reads:

\begin{equation}
\vec{F}_{i} \; ^{'}= \vec{F}_{i} - m_{i} \;  \left[  2 \; \vec{\omega} \times 
\vec{v}_{i} \; ^{'} +  \vec{\omega} \times ( \vec{\omega} \times \vec{r}_{i} \; ^{'}) 
+ \dot{\vec{\omega}} \times \vec{r}_{i} \; ^{'} \right] ,
\end{equation}
where  $\vec{\omega}$ is the rotational frequency of the non-inertial system, i.e.
the Kepler frequency of the corresponding point mass binary, and $\vec{r}_{i}\; ^{'}$ and $\vec{v}_{i}\; ^{'}$ are position and velocity vectors of particle $i$ in the corotating system.
 The first term in the brackets is the Coriolis-force $\vec{F}_{{\rm Cor}}$, the second the centrifugal force $\vec{F}_{{\rm Cent}}$ and the last term, $\vec{F}_{i,\dot{\omega}}$, results from the change of $\vec{\omega}$ due to the gravitational radiation backreaction force. To obtain equilibrium we  apply a velocity dependent damping force, so that the total force on a SPH-particle $i$ reads:

\begin{equation}
\vec{F}^{'}_{i,t}= \vec{F}_{i,{\rm grav}} + \vec{F}_{i,{\rm hydro}} + \vec{F}_{i,{\rm Cent}} + \vec{F}_{i,\dot{\omega}} - m_{i} \gamma \vec{v}^{'}_{i},
\end{equation}
where $\vec{F}_{i,{\rm grav}}$ and $\vec{F}_{i,{\rm hydro}}$ are forces due to the presence of self-gravity and pressure gradients, both of which are evaluated with the SPH-code. We do not have to consider the backreaction force here, since it is applied (in our approximation) to each SPH-particle in the same way and hence does not influence  the shape of the stars. 
The Coriolis force does not have to be considered here since we are interested in the construction of hydrostatic equilibrium where $\vec{v}^{'}_{i}=0$.
As mentioned by Rasio and Shapiro (1992) the relaxation time $\tau= \gamma^{-1}$ should be chosen slightly overcritical to guarantee an efficient relaxation process (the typical oscillation time scale can be estimated from the sound crossing time $\tau_{s}= (G \bar{\rho})^{-\frac{1}{2}}$). \\
We first calculate the resulting forces on the centers of mass and then subtract them from the forces on individual particles in order to remove the force on the stellar centers of mass:

\begin{equation}
\vec{F}^{'j }_{i} \rightarrow \vec{F}^{' j}_{i} - \vec{F}_{cm}^{' j} \label{cmforce},
\end{equation}
where the center of mass forces are given by

\begin{equation}
\vec{F}_{cm}^{' j} = \frac{1}{M_{j}} \sum_{i} m_{i} \vec{F}^{' j}_{i}, \quad \quad j=1,2.
\end{equation}
Here $j$ labels the binary stars, with $M_{j}$ their masses
and the sum loops over the particle numbers in each object.
Because of numerical noise the centers of mass drift very slightly and thus are reset each dump to keep the stars {\em exactly} at the desired positions. \\
Kepler's law $\omega^{2}= G M a_{0}^{-3}$, which we applied in the relaxation process, is strictly valid only for point masses. In their ellipsoidal approximation for polytropic equations of state Lai et al. (1994a) find a modified Kepler law for close equal mass binaries:
\begin{equation}
\omega_{{\rm tidal}}^{2}= \omega^{2} \; (1+2 \delta),
\end{equation}
where
 \begin{equation}
\delta \equiv \frac{3}{2} \; \frac{2 I_{11}-I_{22}-I_{33}}{a_{0}^{2}}
\end{equation}
and  $I_{ii}= \kappa_{n} a_{i}^{2}/5$. The $a_{i}$ are the semi-major axes of the ellipsoid and $\kappa_{n}$ is a constant depending on the polytropic index $n$ ($\kappa_{n}=1$ in the incompressible case, $n=0$; numerical values are tabulated in Lai et al. (1993a)). We calculated $\omega_{{\rm tidal}}$ a posteriori using the parameters of the relaxed system. We found $\omega_{{\rm tidal}} \approx 1.015 \cdot \omega$, which justifies using $\omega$ instead of $\omega_{{\rm tidal}}$ in the relaxation process.\\
Starting with a binary system relaxed in the way described above no oscillations of the stars were observed.

\subsubsection{Viscosity}
Since the neutrons and protons are likely to be in superfluid states, the main contribution to the microscopic viscosity of the neutron star fluid is supposed to come from electron-electron scattering (Flowers $\&$ Itoh 1979). The simulation of the almost inviscid neutron star fluid poses a problem for numerical hydrodynamics since the latter is subject to numerical as well as to artificial viscosity.\\
The SPH standard tensor of artificial viscosity (Monaghan $\&$  Gingold 1983; we use the standard values $\alpha=1$ and $\beta=2$) is known to yield good results in shocks, but also to introduce spurious forces in pure shear flows, where  a vanishing viscosity would be desirable. We thus decided to compare the results obtained using the standard viscosity with those found using a new scheme proposed by Morris $\&$  Monaghan (1997).
The basic idea is to  give each particle its own time dependent viscosity coefficient that is calculated  by integrating a simple differential equation including a source term and a term describing the decay of this coefficient. The viscosity tensor then reads 

 \[ \Pi_{ij} = \left\{ \begin{array}{lr}
\frac{- \alpha_{ij} (t) \; c_{ij} \; \mu_{ij} \;  + \;\beta_{ij} (t)\; \mu_{ij}^{2}}{\rho_{ij}} \quad \quad & \vec{r}_{ij} \vec{v}_{ij} \le 0 \\
\quad \quad \quad  0  & \vec{r}_{ij} \vec{v}_{ij} > 0  ,
\end{array}    \right. \]
where $\beta_{ij}= 2 \; \alpha_{ij}$, $\alpha_{ij}= \frac{\alpha_{i}+\alpha_{j}}{2}$, $c_{ij}= \frac{c_{i} + c_{j}}{2}$, $h_{ij}= \frac{h_{i} + h_{j}}{2}$ and $
\mu_{ij} = \frac{h_{ij}\vec{r}_{ij}\vec{v}_{ij} }{r_{ij}^{2}+\eta h_{ij}^{2}}$. The $c_{k}$ denote the particle sound velocities, $h_{k}$ the smoothing lengths,  $\vec{r}_{k}$ and $\vec{v}_{k}$ positions and velocities,
 $\vec{r}_{ij}=\vec{r}_{i}-\vec{r}_{j}$, $\vec{v}_{ij}=\vec{v}_{i}-\vec{v}_{j}$ and $\eta= 0.01$.
$\alpha_{i}$ is calculated from 

\begin{equation}
\frac{d \alpha_{i}}{dt} = - \frac{\alpha_{i}- \alpha ^{\ast}}{\tau_{i}} + S_{i}.
\end{equation}
The first term on the right hand side  causes the viscosity coefficient $\alpha$  to decay on a  time scale $\tau_{i}$ towards some minimum value $\alpha ^{\ast}$ that keeps the particles well ordered in the absence of shocks. The source term $S_{i}$, leading to an increase of $\alpha_{i}$ when a shock is detected, is chosen to be 

\begin{equation}
S_{i}= max(-( \vec{\nabla} \vec{v})_{i},0).\label{viscsource}
\end{equation}
We chose $\tau_{i}= \frac{h_{i}}{\epsilon \; c_{i}}$ with $\alpha ^{\ast}= \epsilon  = 0.1$.

\subsubsection{Spins}
The degree of synchronization of a close binary system depends decisively on the viscous dissipation rate (compared to the inspiral time). Bildsten $\&$ Cutler (1992), Kochanek (1992) and Lai (1994) found any reasonable assumption on the microscopic viscosity of the neutron star matter to be orders of magnitude too low to lead to a tidal locking of a binary system  within the inspiral time. Hence, it seems very unlikely that a corotating system is the generic case. We suspect the viscosity to be also too low to lead necessarily to an alignment of the stellar spins with the orbital angular momentum. However, for reasons of simplicity we will only treat aligned systems in this work.\\
For purely numerical reasons, i.e. to minimize the effects of viscosity during the inspiral and the shear motion at the contact surface of both stars during the merger, we focus here mainly on synchronized, ``corotating'' systems. This might also be useful as some kind of benchmark for future simulations with a more sophisticated incorporation of general relativistic effects (for a general relativistic simulation of a corotating inspiral see e.g. Baumgarte et al. 1997).\\
However, to explore the dependence of our results on the initial neutron star spins, we also investigate the cases where the stars have no initial spin (in a space-fixed frame) and the case where they are spinning against the orbital motion with a period equal to the orbital period. 
Starting with $45$ km of initial separation, this corresponds to a spin period of $2.91$ ms. We expect all mergers to lie in this range of initial spins.

\subsubsection{Equation of state}
The equation of state is a major uncertainty in all neutron star related calculations. Our code is coupled to the physical equation of state (EOS) for hot and  dense matter of Lattimer and Swesty (1991). This EOS models the hadronic matter as consisting of an average heavy nucleus (representing all heavy nuclei), alpha particles (representative for an ensemble of light nuclei) and nucleons outside nuclei.
 Since further hadronic degrees of freedom, such as hyperons, pions, kaons and the quark-hadron phase transition are disregarded, we use the lowest available nuclear compressibility, $K= 180$ MeV, to mimic the possible softening of the EOS due to the appearance of ``exotic'' matter at higher densities. Since the chemical composition of matter is calculated for the thermodynamically most favourable state, effects of recombinations of nucleons into nuclei are taken into account automatically. For further details concerning the LS-EOS see Appendix B.\\
To test the dependence of our results on the underlying EOS we also use a polytropic EOS where the exponent $\Gamma$ was fitted to the central part of our 1.6 M$_{\odot}$ neutron star obtained with the LS-EOS. We obtained a good fit  with $\Gamma= 2.6$. The pressure $p$ can be written in terms of density $\rho$  and specific internal energy $u_{p}$

\begin{equation}
p= (\Gamma-1) \; \rho \; u_{p}.
\end{equation}
We assign the specific internal energy of our polytrope, $u_{p}$, in a way that the pressure distribution of the LS-EOS star is reproduced:

\begin{equation}
u_{p}(r)= \frac{p_{LS}(r)}{(\Gamma-1) \; \rho(r)},
\end{equation}
where $p_{LS}(r)$ and $\rho(r)$ are the values of pressure and density at radius $r$ in the star. Using these profiles a spherical equilibrium configuration was constructed following the steps described in Appendix A. \\
To test the sensitivity on the adiabatic exponent we also considered a test case of a much softer EOS ($\Gamma= 2.0$) where the polytropic constant $K$ ($p= K \rho^{\Gamma}$) was adjusted in a way that the radius of the LS-EOS star was reproduced.

\subsubsection{Temperatures}

When the neutron stars of a binary system have reached a separation of a few stellar radii (typical time scales are of the order of $10^{8}$ years) they should be essentially cold ($T < 10^{6}$ K). When the stars come closer and tidal interactions  become important, the stars might be heated up again by viscous processes acting in the tidal flow. However, these  temperatures are estimated to be only of the order of $10^{8}$ K (Lai 1994).\\
With respect to the thermodynamic properties, the accurate determination of low temperatures in the high density regime (say above $10^{14} {\rm g cm}^{-3}$) is very difficult. Since the temperature is a very steeply rising function of the specific internal energy $u$ (our independent variables are $u$, the electron fraction $Y_{e}$ and the density $\rho$), numerical noise in $u$ in the degenerate regime may lead to substantial temperature fluctuations. However, as long as there are no additional physical processes involved that depend on temperature, this does not influence the dynamical evolution of matter.

\subsubsection{Neutrino emission}

The merger of two neutron stars will be accompanied by strong neutrino emission (see e.g. Ruffert et al. (1996, 1997)).  To set a limit on the maximum influence  of the cooling by neutrino emission on the ejection of mass, we also explore the extreme case where we assume that all locally produced neutrinos stream out freely without any interaction with overlying material (which also implies that a possible quenching of neutrino rates due to neutrino final-state blocking is ignored), thereby reducing the local thermal energy content.
This scheme is clearly crude and will largely overestimate the emission from the hot, dense central regions. However, we are here only interested in the possible influence on the amount of ejected material and the free streaming limit will give us a {\em secure} upper limit of the effects resulting from neutrino emission.\\
 



Looking at typical densities and temperatures encountered in our scenario, one can conclude that the only processes giving possibly non-negligible contributions to the neutrino emission in addition to  the electron and positron capture on nucleons, are the plasma and the pair processes (for a comparison of the importance of different neutrino processes as a function of $\rho, \; Y_{e}$ and $T$ see e.g. Schinder et al. (1987), Itoh et al. (1989) and Haft et al. (1994)).\\
We thus include the following processes as sources of energy loss:

\begin{eqnarray}
e \; + \; p \; &\rightarrow& \; n \; + \nu_{e} \\
\vspace{0.8 cm}
e^{+} \; + \; n \; & \rightarrow & \; p \; + \bar{\nu}_{e}\\
\vspace{0.8 cm}
e^{+}  \; + \; e \; & \rightarrow & \; \nu_{i} \; + \bar{\nu}_{i}\\
\vspace{0.8 cm}
\gamma_{L,T} \;  & \rightarrow & \;  \nu_{i} \; + \bar{\nu}_{i},
\end{eqnarray}
where the last process is the decay of a plasmon, i.e. the decay of both longitudinal and transverse electromagnetic excitations of the medium. Due to the very high temperatures, we ignore the electron mass $m_{e}$ and the mass difference $Q$ of neutron and proton for our treatment of electron and positron captures. The energy emission rates due to these processes are discussed in Appendix C.
For the energy emission rates of pair and plasma neutrino processes we use the fit formulae of Itoh et al. (1989) and Haft et al. (1994). The neutrino  emission then appears as a sink in the energy equation of each particle:

\begin{equation}
 \frac{du_{i}}{dt} =   \left( \frac{du_{i}}{dt} \right)_{0}  - \frac{Q_{{\rm EC}}+Q_{{\rm PC}}+Q_{{\rm pair}}+Q_{{\rm plas}}}{\rho_{i}},
\end{equation}
where $ \left( \frac{du_{i}}{dt} \right)_{0}$ contains the usual terms from $pdV$-work and viscous heating (see e.g. Benz (1990)).
The enormous dependence 
of the rates on temperature ($Q_{{\rm EC}}, Q_{{\rm PC}} \sim T^{6}$; $Q_{{\rm pair}}, Q_{{\rm plas}} \sim T^{9}$) gave rise to numerical instabilities in the degenerate regime, 
where temperature  fluctuations  can be appreciable due to noise in the specific internal energy.
To overcome this problem the emission rates of particle $i$ were calculated from mean temperatures $\bar{T}_{i}= N_{i}^{-1} \; \sum_{j}^{N_{i}}\; T_{j}$, where $N_{i}$ denotes the number of neighbours of particle $i$, rather than from the individual particle temperatures.

\begin{table*}[tb]
\caption{Summary of the different runs: 1: corotation; 2: no initial spins; 3: spin against the orbit; 
LS: Lattimer and Swesty (1991), P: polytropic EOS  ; MG: Monaghan and Gingold (1983); MM: Morris and Monaghan (1997) }
\begin{center}
\begin{tabular}{cclccccccc} \hline \noalign{\smallskip}
run & spin & EOS  & viscosity & backreaction &  tidal def. & $\nu$ & ${\rm M} \; [ {\rm M}_{\odot}]$ & \#  part. & $a_{0}$ [km] \\ \hline \\
A   & 1 & LS & MG & yes & no  &  no & 1.6 & 20852 & 45 \\ 
B   & 1 & LS & MG & yes & yes &  no & 1.6 & 20852 & 45 \\
C   & 1 & P, $\Gamma= 2.6$  & MG & yes & no  &  no & 1.6 & 20852 & 45 \\ 
D   & 1 & LS & MM & yes & no  &  no & 1.6 & 20852 & 45 \\
E   & 1 & LS & MG & yes & no  &  no & 1.6 & 49974 & 45 \\
F   & 1 & LS & MG & yes & no  &  no & 1.4 & 20852 & 45 \\
G   & 1 & LS & MG & yes & no  & yes & 1.6 & 20852 & 45 \\
H   & 1 & LS & MG & no  & no  &  no & 1.6 & 20852 & 42 \\
I   & 2 & LS & MM & yes & no  &  no & 1.6 & 49974 & 45 \\ 
J   & 3 & LS & MM & yes & no  &  no & 1.6 & 49974 & 45 \\ 
K   & 1 &P, $\Gamma= 2.0$ &  MG & yes & no  &  no  &  1.6 & 20974 & 45 \\ 
\noalign{\smallskip}
\end{tabular}
\end{center}
\label{runs}
\end{table*}

\section{Results}
After having discussed in Sect. 2 the numerical method (including the initial setup) and all possible sensitivities and uncertainties, we have performed in total 11 numerical simulations (run A to K), to test all of these aspects. In this way we intend to analyze  the robustness of the results from the (corotating) standard case and to understand what variation of these results can be expected for different system parameters and physics input. The specifics which characterize each of these runs are given in Table \ref{runs}. The following discussion of the results will refer to these table entries. In all of these runs
we follow the evolution of matter for 12.9 ms.

\subsection{Morphology}
Figs. \ref{morphA1} to \ref{morphJ1} show the particle positions of runs A, C, I and J. The different colors in Fig. \ref{morphA1} label those particles that end up (last dump at $t= 12.9$ ms) in the central object (black), the disk (green), the tails (blue) and those that are unbound at the end of the simulation (red). Run A stands representative for all the corotating runs using the Lattimer/Swesty EOS (the particle plots of runs A, B, D  and E are practically indistinguishable; runs F, G and H differ only slightly).
 The plots only show the first half of the simulation. In the second half practically no additional angular motion is visible, matter just expands  radially up to $\sim 900$ km for all runs apart from the one without initial stellar spins. In this case the outermost layers only extend out to $\sim 600$ km  from the center of mass.\\
In all corotating runs those parts having the largest distance from the common center of mass are driven into thick spiral arms. Those get ``wrapped up'' around the central object to form a ring-like structure (see Figs. \ref{morphA1} and \ref{morphC1}). In the further evolution this ring contracts to form a thick disk, while the matter in the spiral arms continues to wind up around the coalesced, massive object in the center of mass.\\
The corotating runs show, apart from the ``main'' spiral arms, also a very narrow ``side'' spiral structure (see, for example, the second panel in Fig. \ref{morphC1}; in outlines, this structure is also visible in run I), that emerges directly after the stars have come into contact. 
The side spiral arm matter comes from the ``contact zone'' (see panel one in Fig. \ref{morphA1}), is then strongly compressed and finally, in a kind of  ``tube-of-toothpaste''-effect ejected into two narrow jets in opposite directions. While matter in the main arms is smoothly decompressed due to gravitational torques, matter in the side spiral arms gets compressed and heated up.
 The pressure responsible for the formation of the side spiral arms is of thermal nature. This is suggested since no such spiral arms emerge in run G, where neutrino emission leads to a very efficient cooling.  In an additional test run, where the viscous pressure contribution to the energy equation was disregarded, also no side spiral arms occurred.\\
In the end three well-separated structures are visible: an oblate, massive central object, a thick surrounding disk, and long spiral arms. Going to lower total angular momentum, i.e. to runs I and J, the differences between these structures get washed out.\\
It is interesting to note the dependence of the spiral arms on the equation of state: while they are thick and pumped up in all corotating runs using the LS-EOS, they remain narrow and well-defined in the run with the stiff polytropic EOS ($\Gamma= 2.6$, run C; somewhat thicker for $\Gamma= 2.0$, run K). This expansion is a result of an increase of the adiabatic index and the energy release due to the recombination of the nucleons into heavy nuclei.
Fig. \ref{bnuctot} shows the total amount of nuclear binding energy present according to the LS-EOS:
 
\begin{equation}
B_{{\rm nuc,tot}}= \sum_{i} m_{i} [ Y_{h} \; B_{h} + Y_{\alpha} \; B_{\alpha}] \label{bnuc},
\end{equation} 
where the summation ranges over all SPH-particles, $m_{i}$ denotes the particle mass, $Y_{h}$ and $Y_{\alpha}$ the abundances of heavy nuclei and alpha particles, and $B_{h}$ and $B_{\alpha}$ the nuclear binding energies. Around $10^{50}$ erg of nuclear binding energy are deposited (i.e. present in the last dump) in the spiral arms containing only $\sim 0.07 \; {\rm M}_{\odot}$ (see Table \ref{masses}).\\
The degree of conservation of energy and angular momentum can be inferred  from Fig. \ref{conservation}. Both are conserved to approximately $3\cdot10^{-3}$.



\subsection{Mass distribution}
\begin{table*}[th]
\caption{Masses of the different morphological regions, $ \tilde{{\rm M}}_{\rm c.o.}$ is the minimum gravitational mass (see Eq. (22)),
$a$ is the relativistic stability parameter (see text).}
\begin{flushleft}
\begin{tabular}{cccccc} \hline
run & ${\rm M}_{\rm c.o.} \; [{\rm M}_{\odot}]$ & $ \tilde{{\rm M}}_{\rm c.o.} \; [{\rm M}_{\odot}]$  & ${\rm M}_{\rm disk} \; [{\rm M}_{\odot}]$ & ${\rm M}_{\rm tail} \; [{\rm M}_{\odot}]$ & $a$ \\ \hline
A & 2.81 & 2.32 & 0.32 & 0.07 & 0.54\\ 
B & 2.81 & 2.32 & 0.32 & 0.07 & 0.54\\ 
C & 2.78 & 2.29 & 0.35 & 0.08 & 0.51\\ 
D & 2.81 & 2.32 & 0.32 & 0.07 & 0.51\\ 
E & 2.81 & 2.32 & 0.32 & 0.07 & 0.54\\
F & 2.46 & 2.07 & 0.27 & 0.07 & 0.56\\ 
G & 2.88 & 2.36 & 0.25 & 0.07 & 0.58\\ 
H & 2.75 & 2.27 & 0.32 & 0.12 & 0.49\\
I & 3.05 & 2.47 & 0.11 & 0.04 & 0.62\\ 
J & 3.14 & 2.54 & 0.06 & $8\cdot 10^{-3}$ & 0.61\\ 
K & 2.76 & 2.28 & 0.40 & 0.04 & 0.53\\
\end{tabular}
\end{flushleft}
\label{masses}
\end{table*}
Density contours (cuts through the y-z-plane at the last dump, t=12.9 ms) of run E (representative for the corotating runs), run G (to see the influence of the cooling by neutrino emission), run I and run J (to see the influence of the initial spins) are shown in Figs. \ref{yzbw1} and \ref{yzbw2}.
The distribution of mass with density and radius is shown in Figs. \ref{mvsrho} and \ref{mvsr}. The sharp bends in  Fig. \ref{mvsrho} indicate the separations of the different morphological regions.
The amounts of mass in the central object, the disk and the tails (collectively used for the low-density material outside the disk) can be inferred from this plot, the corresponding masses are listed in Table \ref{masses}.
The runs A, B, D and E practically coincide in Fig. \ref{mvsrho}, suggesting 
that the mass distribution is insensitive to our approximation of initially 
spherical stars, to the details of artificial viscosity, and also shows that 
the resolution in our standard run is sufficient to describe the mass 
distribution properly.
The temporal evolution of the masses and densities in the three regions 
(run A) can be inferred from Fig. \ref{mvsrhovst} (Fig. \ref{mvsrho} just 
corresponds to the last temporal slice of Fig. \ref{mvsrhovst}). In the 
beginning practically all the mass ($= 3.2 \; {\rm M}_{\odot}$) has a density 
above $10^{14}{\rm g cm}^{-3}$ (the visible oscillations result from  the 
initially spherical stars). Around t$= 2.5$ ms a strong expansion sets in 
(less mass has a density above, say, $10^{14}{\rm g cm}^{-3}$). Then, around 
t$= 4.5$ ms, the three regions (central object, disk and tails) become visible.
 At around t$= 5.5$ ms, when the wrapped-up spiral arms are contracted by 
gravity, a recompactification corresponding, to the ``bump'' in Fig. 
\ref{mvsrhovst} sets in (to localize it in time and density the contour lines 
of 2.8, 2.9, 3.0 and 3.1 ${\rm M}_{\odot}$ are projected on the 
log($\rho$)-t-plane). This recompactification is accompanied by  a reheating 
(this is also reflected in the neutrino emission of the disk, where a strong 
 increase is visible, compare panel 2 in Fig. \ref{nunucmult}) that finally 
reexpands the disk. When our simulation stops, the disk has a density range 
from $ \sim 10^{10} - 10^{12} \; 
{\rm g cm}^{-3}$ (compare also to the contour plots \ref{yzbw1} and 
\ref{yzbw2}), densities above this range are associated with the central 
object, densities below with the tails.\\
The overall mass distribution is shown  in Figs. \ref{xyEbw} and \ref{xyIbw}.

\subsubsection{The central object}
Figs. \ref{xyECbw} and \ref{xyIJbw} show density contours of runs E, C, I and J.
The central objects (especially for runs A to H) consist of a core with densities above $10^{14} {\rm g cm}^{-3}$  and a diffuse edge ranging from $10^{14}$ to $\sim 10^{12.5}{\rm g cm}^{-3}$  forming a kind of a "hot skin" around the core (see the smooth transition from the central object to the disk in Fig. \ref{mvsrho}). 
In the run including neutrino energy losses (run G) there is a very sharp transition corresponding to a huge density gradient from the central object to the disk, suggesting that the edge diffuseness of the central object in the other runs emerges from thermal pressure (also visible in the temperature, Fig. \ref{TmixE}). Because of the absence of this pressure, matter of the order  $0.07 \; {\rm M}_{\odot}$, otherwise located in the disk, has fallen onto the central object increasing its mass to $2.88 \; {\rm M}_{\odot}$. (This hot skin might be an artifact of the viscosity scheme, since it less pronounced in run D).
The masses of the final central objects range from $\sim 2.5 \; {\rm M}_{\odot}$ to $\sim 3.1 \; {\rm M}_{\odot}$ (see Table \ref{masses}) and are thus well above the maximum precisely known neutron star  (gravitational) mass of $1.44 \; {\rm M}_{\odot}$ for the pulsar of the binary system PSR 1913+16 (Taylor 1994; there are, however, objects, whose mass error bars reach up to 2.5 ${\rm M}_{\odot}$; see Prakash (1997b) and references therein). Theoretical calculations for maximally rotating neutron stars, using a large set of 17 (11 relativistic 
nuclear field theoretical and 6 non-relativistic potential models for the nucleon-nucleon-force) nuclear equations of state (Weber $\&$ Glendenning 1992) find maximum values for the gravitational masses of $\sim 2.5 \; {\rm M}_{\odot}$. Since these values refer to gravitational rather than to baryonic masses (which we are referring to due to the Newtonian character of our calculations) we have to estimate the {\em gravitational} masses of our central objects. The relation obtained by Lattimer and Yahil (1989) for reasonable uncertainties in the equation of state reads

\begin{equation}
E_{\rm bin}= (1.5 \pm 0.15) \cdot \left( \frac{{\rm M}_{g}}{{\rm M}_{\odot}} \right) ^{2} \cdot 10^{53} \; {\rm erg}.
\end{equation}
This gives the minimal gravitational mass $ \tilde{\rm M}_{g} $ of the baryonic mass ${\rm M}_{b}$ as

\begin{equation}
\frac{\tilde{\rm M}_{g}}{{\rm M}_{\odot}}= 5.420 \cdot [-1 + \sqrt{1+0.369 \; \frac{\rm M_{b}}{{\rm M}_{\odot}}}],
\end{equation}
which is typically 0.4 - 0.5 ${\rm M}_{\odot}$ smaller than the baryonic mass.
The corresponding values for our central objects are shown as entry 3 in Table \ref{masses}). These values are still very high, but there are equations of state  for which fast rotating neutron stars in this mass range are stable. Thermal pressure, which is disregarded in the work of Weber and Glendenning since they assume old neutron stars, could further stabilize the central object on a cooling time scale.\\
The above quoted masses refer to neutron stars rotating with the maximum possible velocity. To estimate the influence of rotation for our calculations we plot in Fig. \ref{velcomp} the tangential velocities in the central objects of run E, I and J and compare them to the Kepler velocity $v_{K}= (G M(r)/r)^{1/2}$, i.e. the velocity where the centrifugal forces balance the gravitational forces ($M(r)$ is the mass enclosed in the cylindrical radius r). In all cases the velocities are clearly below the Kepler velocity and we thus do not suppose rotation to play an important role for stabilization (in this Newtonian consideration). In their general relativistic  analysis Weber and Glendenning find  the maximum possible rotation frequency $\Omega_{K} \sim 1.2 \cdot 10^{4} \; {\rm s}^{-1}$ for the equations of state referred to above corresponding to a maximum stable mass of $M \approx 2.5 \;  {\rm M_{\odot}}$ (Weber $\&$ Weigel 1989; Weber et al. 1991). This is larger than our rotation frequencies by a factor of about two.\\
In the general relativistic case the stability support from rotation is determined by the dimensionless parameter $a= \frac{J c}{G M^{2}}$ (Stark $\&$ Piran 1985). We give $a$ for our central objects in  column six in Table \ref{masses}. We find values of $\approx 0.5$ in agreement with previous simulations (Shibata et al. 1992, Rasio  $\&$  Shapiro 1992, Ruffert et al. 1996). This is well below the critical value $a_{crit} \approx 1$, meaning that the central object cannot be stabilized against collapse by rotation.\\
Prakash et al. (1995) studied the influence of the composition on the maximum neutron star mass. Their general result is that trapped neutrinos together with nonleptonic negative charges (such as $\Sigma^{-}$ hyperons, or d and s quarks) lead to an increase of the maximum possible mass (in contrast to nucleons-only matter). Thus, the collapse of a star with almost the maximum mass could be delayed on a neutrino diffusion time scale. As an estimate of this time scale for our central objects, we use R $ \approx 15$ km (center to pole distance, see Fig. \ref{yzbw1}) and a typical neutrino energy of E~$_{\nu} \approx \langle {\rm E}_{\nu_{e}} \rangle \approx {\rm T \; \frac{F_{5}(0)}{F_{4}(0)}} \approx 50$ MeV, where T$ \approx 10 $ MeV and $ {\rm \frac{F_{5}(0)}{F_{4}(0)}} \approx 5$ has been used. This gives a mean free path of 

\begin{equation}
\lambda \approx \frac{1}{n \sigma} \approx 0.75 \; {\rm m},
\end{equation}
where the baryon number density n corresponding to $5\cdot 10^{14} {\rm g cm^{-3}}$ and a neutrino nucleon scattering cross section  $\sigma = \frac{1}{4}  \sigma_{0}  \left( \frac{E_{\nu}}{m_{e} c^{2}} \right) ^{2}$ with $\sigma_{0} = 1.76\cdot 10^{-44}$ cm$^{2}$ (see Tubbs $\&$ Schramm 1975) has been used. An estimate for the delay time scale is then given by 
\begin{equation}
\tau \approx \frac{3 {\rm R}^{2}}{\lambda c} \approx 3 \;  {\rm s} \label{taunu}.
\end{equation}
Thus if there were nonleptonic negative charges present in the central object, the collapse could be delayed for a few seconds.\\
To summarize the stability question of the central object: if all the stabilizing effects mentioned above (EOS, rotation, thermal pressure and exotic matter) should conspire, the central object could (at least in some cases) be stabilized  against gravitational collapse. However, we regard this possibility to be fairly unlikely.\\
If the central object collapses to a black hole, the question arises of how much mass has enough angular momentum to avoid being swallowed. For a simple estimate we assume that the specific angular momentum of a particle  must be larger than the one of a test particle with Kepler velocity at the marginally stable circular direct orbit of a Kerr black hole (see Bardeen et al. 1972).
Thus, $l_{i}= |\vec{r}_{i} \times \vec{v}_{i}| > R_{{\rm ms}} v_{K}(R_{{\rm ms}})= (G M_{{\rm c.o.}} R_{{\rm ms}})^{-1/2}$, where 
\begin{equation}
R_{{\rm ms}}= \frac{G}{c^{2}} M_{{\rm c.o.}} \; \left \{ 3 + z_{2} - [ (3-z_{1})(3+z_{1}+2 z_{2}) ] ^{1/2} \right \}  
\end{equation}
with 
\begin{eqnarray}
z_{1}= &&1 + \left (1-\frac{l_{{\rm c.o.}}^{2}}{M_{{\rm c.o.}}^{2}} \right )^{1/3} \left [ \left (1+\frac{l_{{\rm c.o.}}}{M_{{\rm c.o.}}} \right )^{1/3} + \right. \nonumber \\ && \left. \left ( 1-\frac{l_{{\rm c.o.}}}{M_{{\rm c.o.}}} \right )^{1/3} \right ]
\end{eqnarray}
and
\begin{equation}
z_{2}= \left (\frac{3 l_{{\rm c.o.}}^{2}}{M_{{\rm c.o.}}^{2}} + z_{1} ^{2} \right )^{1/2}.
\end{equation}
$l_{{\rm c.o.}}$ denotes the specific angular momentum of the central object.
 With this estimate we find that the central black hole will be surrounded by a disk of $\sim 0.3$ M$_{\odot}$ for initial corotation ($\sim$ 0.1 and $\sim$ 0.06 for the cases of no initial stellar spins and spins against the orbit). 


\subsubsection[The disk]{The disk}

In all runs the central object is finally surrounded  by a thick  disk (see 
Figs. \ref{yzbw1} and \ref{yzbw2}). In the case of initial corotation the disks contain around  $0.3 \; {\rm M}_{\odot}$ ($0.27 \; {\rm M}_{\odot}$ in run F; 0.25 in run G, where $0.07 \; {\rm M}_{\odot}$ have fallen  onto the central object; in run H, neglecting the backreaction force leads to an impact with a higher angular momentum and thus  more mass is expelled into the tails; the disk, however, also contains $0.3 \; {\rm M}_{\odot}$).
 In all cases, apart from run J,  empty funnels form above the poles of the central object. These are strongly enlarged in run G, where the efficient cooling by neutrinos leads to strongly reduced thermal pressure contributions, i.e. to a softening of the EOS that makes matter more prone to the centrifugal forces thereby leading to a flattening of the central object as well as of the disk. These funnels have the appealing feature that here large gradients of radiation pressure can be built up which can lead to two well-collimated jets in opposite directions, pointing away from the poles. 
These funnels would be an ideal site for a fireball scenario (Goodman 1986; Shemi $\&$ Piran 1990; Paczy$\acute{{\rm n}}$ski 1990; Piran $\&$ Shemi 1993) since they are practically free of baryons. A baryonic load as small as $10^{-5}\; {\rm M}_{\odot}$ could prevent the formation of a GRB. Our present resolution (the lightest SPH-particle has a mass of a few times $10^{-5}$ M$_{\odot}$) is presently still too low to draw conclusions on this point.

\subsubsection[Tails]{The tails}
Despite considerable morphological differences -exploding tails with the Lattimer/Swesty EOS, well-defined, narrow tails for the polytrope- the amount of mass in the tails is rather insensitive to the EOS (see Table \ref{masses}). It is mainly dependent on the total amount of angular momentum during the impact, thus leading to the most massive tails in run H (no angular momentum lost in gravitational waves) with a decreasing tendency going from the standard run to runs I and  J.

\subsection{Temperatures and vortex structures}
\begin{table*}
\caption{Kinetic and thermal energies (in erg) in the different morphological regions.}
\begin{flushleft}
\begin{tabular}[th]{ccccccc} \hline 
run & $ E_{{\rm kin, \; c.o.}}$ & $E_{{\rm kin, \; disk}}$ & $E_{{\rm kin, \; tail}}$ & $E_{{\rm therm,  \;c.o.}}$ & $E_{{\rm therm, \; disk}}$ & $E_{{\rm therm, \; tail}}$\\ \hline \\
A & $1.0 \cdot 10^{53}$ & $ 1.6 \cdot 10^{52}$ & $ 1.4 \cdot 10^{51}$ & $ 2.3 \cdot 10^{52}$ & $ 4.6 \cdot 10^{51}$ & $ 3.9 \cdot 10^{49}$ \\ 
B &$ 9.9 \cdot 10^{52} $&$ 1.5 \cdot 10^{52} $&$  1.5 \cdot 10^{51}$ & $ 2.2 \cdot 10^{52}$ & $ 4.1 \cdot 10^{51} $&$ 3.3 \cdot 10^{49}$\\ 
C &$ 8.3 \cdot 10^{52} $&$ 1.6 \cdot 10^{52} $&$ 1.1 \cdot 10^{51} $&$ 1.3 \cdot 10^{52} $&$ 1.9 \cdot 10^{51}$&$ 4.5 \cdot 10^{49}$\\ 
D &$ 1.0 \cdot 10^{53} $&$ 1.7 \cdot 10^{52} $&$ 1.5 \cdot 10^{51} $&$ 2.3 \cdot 10^{52} $&$4.6 \cdot 10^{51} $&$ 3.8 \cdot 10^{49}$\\ 
E &$ 1.0 \cdot 10^{53} $&$ 1.8 \cdot 10^{52} $&$ 1.5 \cdot 10^{51} $&$ 2.3 \cdot 10^{52} $&$4.4 \cdot 10^{51} $&$ 4.0 \cdot 10^{49}$\\ 
F &$ 7.6 \cdot 10^{52} $&$ 1.2 \cdot 10^{52} $&$ 1.5 \cdot 10^{51} $&$ 1.8 \cdot 10^{52} $&$ 3.5 \cdot 10^{51}$&$ 3.9 \cdot 10^{49}$\\ 
G &$ 1.2 \cdot 10^{53} $&$ 1.4 \cdot 10^{52} $&$ 1.4 \cdot 10^{51} $&$ 1.1 \cdot 10^{52} $&$1.9 \cdot 10^{51} $&$ 3.5 \cdot 10^{49}$\\ 
H &$ 8.5 \cdot 10^{52} $&$ 1.6 \cdot 10^{52} $&$ 2.7 \cdot 10^{51} $&$ 2.1 \cdot 10^{52} $&$ 4.8 \cdot 10^{51}$ & $ 2.2 \cdot 10^{50}$ \\ 
I &$ 1.3 \cdot 10^{53} $&$ 5.1 \cdot 10^{51} $&$ 3.4 \cdot 10^{50} $&$ 4.2 \cdot 10^{52} $&$1.9 \cdot 10^{51} $&$ 1.3 \cdot 10^{49}$\\ 
J &$ 1.4 \cdot 10^{53} $&$ 1.9 \cdot 10^{51} $&$ 2.1 \cdot 10^{50} $&$ 7.7 \cdot 10^{52}$&$ 1.1 \cdot 10^{51} $&$ 2.7 \cdot 10^{49}$\\ 
K &$9.1 \cdot 10^{52} $  & $2.0 \cdot 10^{52} $ & $4.7 \cdot 10^{50} $ & $2.1 \cdot 10^{53} $ & $3.5 \cdot 10^{51} $ & $5.0 \cdot 10^{49} $ \\

\end{tabular}
\end{flushleft}
\label{energies}
\end{table*}
Figure \ref{Tmax} shows the maximum temperatures during the different runs. The curves in the upper panel refer to the maximum temperature of a single particle, the ones below to the maximum smoothed temperature $\langle T \rangle_{i}$, 

\begin{equation}
\langle T \rangle_{i}= \sum_{j} \frac{m_{j}}{\rho_{j}} T_{j} W(|\vec{r_{i}} -\vec{r_{j}}|,h_{ij}),
\end{equation}
where $T_{j}$ are the temperatures, $m_{j}$ the masses, $\rho_{j}$ the densities, $W$ the spherical spline kernel (see e.g. Benz 1990) and $h_{ij}$ the arithmetic mean of the smoothing lengths of particles $i$ and $j$.\\
The smoothed (the particle) temperatures reach peak values of about 50 (80) MeV for run J where the most violent shear motion is present, around 45 (70) MeV in run I and about 30 (50) MeV in the corotating runs.
The resolution and viscosity seem to be of minor importance for the temperature calculation. Starting with spherical stars leads to an increased temperature since oscillation energy is transformed into heat (see also Table \ref{energies}).
In the corotating runs the hot band that forms at the contact surface dissolves into two hot spots (see Fig. \ref{TmixE}). This structure develops further to finally form a hot, s-shaped band through the central object.
For an understanding of these structures we plotted in the right columns of Figs. \ref{TmixE} to \ref{TmixJ} the projections of those particles that are contained within a thin slice ($|z|<0.5$ km). The projected particle positions of star one and two are marked with different symbols. One sees the formation of two macroscopic vortices that can be identified with the  hot spots. The panels in the second line of Fig. \ref{TmixE} show patterns that are typical for Kelvin-Helmholtz instabilities (see e.g. Drazin and Reid 1981). The basic properties of this process, the formation of a hot band in the contact region, separation into two (hot) vortices and the final s-shaped hot band, are unaffected by resolution and the change from the standard SPH-viscosity to the new scheme. The finger like structures that extend into the matter of each star (last panel Fig. \ref{TmixE}) are just broader with lower resolution (run A). With the new viscosity scheme substructures form along the ``fingers'' leading to a more fractal appearance of the line separating the material of both stars. This is a result of a lower viscosity which leads to a dissipation of the energy of turbulent eddies on smaller scales  $\lambda_{v} \approx L R^{-4/3}$ (see e.g. Padmanabhan (1996)), where $L$ is is a macroscopic scale and $R$ the Reynolds number.


In run I three macroscopic vortices, a large central one and two smaller ones to the left and right, form along the contact surface. The two smaller vortices get attracted by the central one and fuse on a time scale of approximately one millisecond. Approximately eight milliseconds after impact the material of both stars is well mixed and wrapped up around the central vortex.\\ 
In the case with maximum shear motion, run J, we  find one large vortex in the centre and several smaller ones along the contact surface. The smaller ones on each side of the center merge so that there are in total three such macroscopic vortices. The outer vortices move towards the origin and finally merge with the central one. The material of both stars gets mixed turbulently on a  time scale of a few milliseconds.\\
Ruffert et al. (1996) also find two macroscopic vortices for the corotating case. However, they find in their simulation that the vortices dissolve practically independently of the initial spin state into a ringlike structure, while our hot spots develop into an s-like shaped hot band along the line separating the matter of the different stars. Since Ruffert et al. do not find a particular influence of the shear motion on the growth time scale and since their results for different initial spins look very similar, they question on the Kelvin-Helmholtz picture and propose an alternative, macroscopic explanation for the flow pattern. 
A quantitative analysis of the incompressible, inviscid case  in terms of linear normal mode analysis
yields a growth rate for an unstable mode of wavelength $\lambda$ (for the case $\rho_{1}=\rho_{2}$; see e.g. Padmanabhan (1996))

\begin{equation}
\sigma \equiv Im(\omega)^{-1} = \frac{\lambda}{\pi v}.
\end{equation}
This implies that the shortest wavelengths will grow fastest and that the perturbation should grow faster with larger shear velocity.\\
The shortest wavelengths to grow are determined by our numerical resolution. Let us assume that this length scale corresponds to the typical distance over which neighbours can interact, i.e. $2 h_{i}$, where the smoothing length $h_{i}$ is to be taken in the shear region. Then we find $\lambda \sim 2 h_{i} \sim 5$ km. We then look at the shear velocities $v$. Here we find by looking at relative velocity projections along the shear interface values of $\sim 0.1 c$ for the corotation, $\sim 0.3 c$ for run I and $\sim 0.5 c$ for run J. Thus typical growth time scales in the different runs should be $\sigma_{A} \sim 5 \cdot 10^{-5}$ s,  $\sigma_{I} \sim 2 \cdot 10^{-5}$ s  and $\sigma_{J} \sim 1 \cdot 10^{-5}$ s, where the subscripts label the runs ($A$ is representative for corotation). This means that the perturbations  have enough time to grow into the macroscopic regime on a dynamical time scale of the system (milliseconds). The growth times scale approximately like 1 : 2 : 5 and this is approximately what is seen in our simulations.
Since in our calculations a strong dependence of the growth time scale on the shear motion, consistent with the Kelvin-Helmholtz time scales, is encountered and the vortex structures for different initial spins are clearly different, we  do not see the necessity of an alternative explanation.


\subsection{Neutrino emission}

The main reason for an additional run with a very simple inclusion of neutrinos is to investigate whether neutrino emission has a noticeable effect on the amount of ejected mass. 
To estimate the appropriateness of our neutrino treatment for the different regions we estimate typical neutrino diffusion time scales. A typical neutrino diffusion time scale for the central object  
is  of the order of a few seconds (see discussion above). Thus our neutrino treatment  is definitely inappropriate there (typical time scales are milliseconds). Applying the same formula (\ref{taunu}) for the disk, using $E_{\nu}=$ 10 MeV, $\rho= 10^{11} {\rm g cm^{-3}}$ and $R=$ 60 km, we find $\tau_{\nu,{\rm disk}} \approx 4 \cdot 10^{-4}$ s, which is comparable to the dynamical time scales. Thus at least in the outer regions of the disk, the free streaming approximation might  be justified. In the tails free streaming neutrinos are clearly a good approximation (apart from, perhaps, the very first moments after impact).
In Fig. \ref{nunucmult} we compare the amount of nuclear binding energy present (see Eq. (\ref{bnuc})) to the amount  of energy radiated in neutrinos
\begin{equation}
E_{\nu}(t) = \int_{0}^{t}(L_{{\rm EC}}(t')+L_{{\rm PC}}(t')+L_{{\rm pair}}(t')+L_{{\rm plas}}(t'))dt',
\end{equation}
where  $L_{\lambda}$ denotes the neutrino luminosity of process $\lambda$.
The $L_{\lambda}$ are given by 
\begin{equation}
L_{\lambda}= \sum_{i} \frac{Q_{\lambda,i}(t)}{\rho_{i}(t)} \; m_{i},
\end{equation}
where $Q_{\lambda,i}$ is the energy emission rate of process $\lambda$ and particle $i$ (in erg s$^{-1}$cm$^{-3}$). Until the end of the simulation the energy lost in neutrinos exceeds that gained from nuclear processes by two orders of magnitude. This is mainly due to the central object, where unphysical amounts of neutrinos are emitted. In reality we expect the disk to be the dominant source  of neutrino emission. When the disk reaches the above mentioned reheating phase the neutrino emission exceeds the nuclear energy by about one order of magnitude (see panel two in Fig. \ref{nunucmult}.)\\
The most important result is that in the spiral arms more energy is gained by nuclear processes  than is lost by neutrino emission (in spite of our overestimate). Such conclusions were already reached in Davies et al. (1994) for low density regions. This is a crucial point since it shows  that  our results concerning mass ejection are independent of the neutrino treatment.

\subsection{Ejected mass and nucleosynthesis}

We regard a particle to be unbound if the sum of its energies -macroscopic as well as  microscopic- is positive, i.e. if

\begin{equation}
E_{{\rm pot}}+E_{{\rm kin}}+E_{{\rm int}} > 0.
\end{equation}
For $E_{int}$ we count all kinds of internal energies apart from the (negative) nuclear binding energies, i.e.

\begin{equation}
E_{{\rm int}} = E_{\gamma} + E_{e} + E_{N},
\end{equation}
where the indices $\gamma, \; e$ and $N$ denote photons, electrons and nucleons. All these terms are taken from the Lattimer/Swesty EOS (for the polytropic case we just use $E_{{\rm int}}= u_{i} m_{i}$). We do not consider nuclear binding energies, since we regard an isolated nucleus at $T= 0$ (the only 
internal energy comes from nuclear binding) with $E_{pot}+E_{kin} > 0$ as unbound. However, near the end of the evolution $E_{{\rm int}}$ is negligible and does practically not influence the total amount of ejected material.\\
This criterion can be cross-checked by adopting a simple model, where we regard the particles as free point masses, i.e. we neglect hydrodynamic forces resulting from pressure gradients and disregard internal degrees of freedom (internal energies). The particles are supposed to  move on Kepler orbits around a point mass in the origin with $M=M_{{\rm c.o.}}+M_{{\rm disk}}$ ($M_{{\rm c.o.}}$ is the mass of the central object, see Table \ref{masses}), which is large compared to the particle masses $m_{i}$, $ \frac{m_{i}}{M} \ll 1$. Under these assumptions the numerical eccentricities of the orbits are given by
\begin{equation}
e_{i}= \sqrt{1+ \frac{2 E_{i} J_{i}^{2}}{G^{2} m_{i}^{3} M^{2}}}, \label{e}
\end{equation}
where $E_{i}$ is the sum of the particle's kinetic and potential energy and $J_{i}$ its angular momentum. We generally find a good agreement of both criteria, with deviations lying in the range of $1 \%$.\\
\begin{table}
\caption{Amount of mass that is unbound at the end of the simulation}
\begin{flushleft}
\begin{tabular}[b]{ccc} \hline 
run & remark & m$_{{\rm esc}}$ $ [ {\rm M}_{\odot} ] \;$ ( \# part.) \\ \hline\\
A & 'standard' (see text)  & $3.1 \cdot 10^{-2}\;$ (295) \\
B & tidally deformed stars & $3.4 \cdot 10^{-2}\;$ (324)\\ 
C & polytropic EOS, $\Gamma= 2.6$         & $1.5 \cdot 10^{-2}\;$ (131)\\ 
D & lower viscosity        & $3.3 \cdot 10^{-2}\;$ (318)\\ 
E & double particle number & $3.2 \cdot 10^{-2}\;$ (844)\\
F & $1.4 \;{\rm M}_{\odot}$& $3.6 \cdot 10^{-2}\;$ (373) \\ 
G & neutrinos              & $3.1 \cdot 10^{-2}\;$ (291) \\ 
H & no backreaction        & $5.4 \cdot 10^{-2}\;$ (523) \\ 
I & no spins               & $8.7 \cdot 10^{-3}\;$ (208) \\ 
J & spins against orbit    & $5.2 \cdot 10^{-3}\;$ (130) \\ 
K &polytropic EOS, $\Gamma= 2.0$ & $ 0 $ (0) \\
\end{tabular}
\end{flushleft}
\label{ejecta}
\end{table}

All corotating runs using the LS-EOS eject around $3 \cdot 10^{-2} \; {\rm M}_{\odot} $, about twice as much as the run using the stiff polytrope. This is caused by a variation of the adiabatic exponent and the formation of nuclei when matter is decompressed and thereby releases the gained nuclear binding energy. For reasons of illustration we plot in Fig. \ref{mesct}  the amount of unbound mass versus the time and compare this with the amount of nuclear binding energy present in the mass that ultimately escapes (see Fig. \ref{bnucesc}).
The flatness of the curves in Fig. \ref{mesct} indicates that not much more mass will be ejected during the further evolution. The rise in the adiabatic exponent and the deposition of a few times $10^{49}$ erg in $\sim 3 \cdot 10^{-2} \;  {\rm M}_{\odot} $ (see Table \ref{ejecta}) leads to an explosive expansion  of the spiral arm tips, thereby supporting the ejection of mass.
Tidal deformation (run B) leads to an increase of the ejected mass since the system contains more angular momentum due to its elongated shape. As expected, the 1.4 ${\rm M}_{\odot} $ run ejects more mass since the gravitational potential to be overcome is shallower than in the 1.6 ${\rm M}_{\odot} $ case. We suspect this to be true also in the general relativistic case where the less massive stars have larger radii and their outer parts therefore contain more angular momentum. 
The realistic reduction of viscosity also tends to increase the amount of ejecta. The inclusion of neutrinos does not alter the  results concerning unbound mass (the emitted energy in neutrinos is approximately one order of magnitude lower than the released nuclear binding energy for the ejected matter, see panel three in Fig. \ref{nunucmult}). The basic intention of run H was to test the sensitivity of the amount of ejected mass on the details of the treatment of the gravitational radiation backreaction force. Here almost twice as much matter is ejected (since no angular momentum is lost in gravitational waves), indicating that our results could be influenced by our simplified treatment of this force (see Eqs. (\ref{gwback1}) and (\ref{gwback2})). In the runs with lower initial angular momentum (run I and J) the amount of ejecta is substantially 
lower, indicating a very strong dependence on the initial spins. The sensitivity to the EOS is underlined by the fact that in the run with the soft polytrope ($\Gamma= 2.0$) no resolvable amount of mass is ejected, in agreement with the result of Rasio and Shapiro (1992).\\
The amount of r-process material that could be formed in this merging scenario is basically determined by $Y_{e}$ and the entropies (and expansion time scales). In Fig. \ref{escprop} we plot entropies (in $k_{B}$ per nucleon) and densities at the time of ejection for the three different neutron star spins (run E, I and J). It is very interesting to note that the densities at the moment of ejection are very high, the bulk of matter becomes unbound at densities from $\sim 10^{14}$ to $\sim 10^{12}$ g cm$^{-3}$. This is well above the neutron drip ($4 \cdot 10^{11}$ g cm$^{-3}$), where in spite of the high temperatures very large ($A \sim 250$ according to the LS-EOS) and very neutron rich ($Z/A \sim 0.15$) nuclei are present in appreciable amounts ($X_{h} \sim 0.3$). These nuclei are far from being experimentally well-known. Hence, to start r-process calculations from these initial conditions, very exotic nuclei (not in vacuo, but immersed in a dense neutron gas) have to be implemented  in the corresponding reaction networks.
In addition, the effects of the high Fermi-energies on the reaction rates including beta decays (Pauli-blocking) have to be accounted for.\\
The temperatures are strongly dependent on the ejection mechanism which is closely related to the stellar spins. In the corotating runs the ejecta are initially located on the front side of each star (with respect to the orbital velocity). Due to gravitational torques this matter gets smoothly stretched and thereby decompressed, no spikes in pressure, temperature and the time dependent viscosity parameter $\alpha$ (see Eq. \ref{viscsource}) are visible. In this case the temperatures are only slightly above the initial temperature ($4 \cdot 10^{9}$ K; see Appendix A). We suspect this temperature increase to be mainly due to (artificially high) viscosity. 
The material is ejected in a different way for the other spin configurations.
In the case without stellar spins the ejecta can be separated into two groups according to their ejection mechanism. The material of the first group is found in the spiral arm structure, ejected in a way similar to corotation and is thus essentially cold. The second group comes from a region that gets strongly compressed and thereby heated up to temperatures
around 6 MeV. In the following expansion, however, this material cools down quickly. In the case where the stars spin against their orbital motion all the material is ejected by the second mechanism thereby  reaching even higher temperatures in the compression phase (around 9 MeV) for a short time. \\
Since the material that gets unbound in the coalescence of initially corotating systems stays essentially cold ($10^{8}$ K, see Lai 1994), we expect the $Y_{e}$ of this matter to be close to the initial values of the cold neutron stars, i.e. 0.01 $\le Y_{e} \le$ 0.05 with small contributions from the stellar crust. The cases with different stellar spins eject material that gets heated appreciably before being cooled by the expansion.
In these cases $Y_{e}$ might be different from the initial values, since temperatures are high enough for the charged current reactions ($e^{-}-,e^{+}-$capture) to set in at non-negligible rates. \\
Owing to the problems in explaining the observed r-process abundances entirely by type II supernovae, there seems to be a need for at least one further astrophysical scenario that is able to produce  r-process nuclei in appreciable amounts. Neutron star mergers  are attractive candidates since they would in a natural way provide large neutron fluxes, low $Y_{e}$s and moderate entropies (which provides r-process matter more easily than high $Y_{e}$ and entropy conditions). An r-process under such conditions should be very efficient and produce mostly elements in the high mass region. Thus, perhaps all of the r-process matter with $A > 110$, that can only be produced in the right amounts in supernova calculations if artificially high entropies are applied (see Freiburghaus et al. 1997; Takahashi et al. 1994), perhaps all of this matter could be synthesized in neutron star binary (or BH-NS) mergers.\\
Assuming a core collapse supernova rate of $2.2 \cdot 10^{-2}$ (year galaxy)$^{-1}$ (Ratnatunga 1989), one needs $10^{-6}$ to $10^{-4} \; {\rm M}_{\odot} $ of ejected r-process material per supernova event to explain the observations  if type II supernovae are assumed to be the only source. The rate of neutron star mergers, which is by far more uncertain, has recently been estimated to be $8 \cdot 10^{-6}$ (year galaxy)$^{-1}$ (see van den Heuvel $\&$ Lorimer 1996). Taking these numbers, one would hence need $\sim 3 \cdot 10^{-3} \; {\rm M}_{\odot}$ to $\sim 0.3 \; {\rm M}_{\odot}$ per event for an explanation of the observed r-process material  exclusively by neutron star mergers. Thus our  results for the ejected mass from $4 \cdot 10^{-3}$ to $3-4 \cdot 10^{-2}  \; {\rm M}_{\odot}$ look promising (see Fig. \ref{ratecomp}).
Meyer (1989) calculated the decompression of initially cold neutron star 
material ($T= 0$ K) assuming  the expansion to be given by multiples of the 
free-fall time scale. He found that the decompressed neutron star material 
gives always, i.e. regardless of the expansion rate, rise to r-process 
conditions. Thus even the initially cold ejecta from corotating 
configurations should heat up during the expansion and form r-process nuclei.
If, as suggested by Meyer (1989), large parts of the ejected material should 
consist of r-process nuclei, neutron star mergers could account for the whole 
observed r-process material in the galaxy. 
However, whether the observed abundance patterns can be explained with this 
scenario remains an open question and is left to further investigations.

\section{Summary}
We have presented the results of three-dimensional Newtonian hydrodynamic simulations of the merger of equal mass neutron star binary systems. We followed the evolution of two neutron stars containing 1.6 (1.4) M$_{\odot}$ of baryonic matter starting with an initial center of mass distance of 45 km for $\sim$ 13 ms. In a total of 11 runs we tested  the sensitivity of the results to the physical parameters of the binary system and to a number of model assumptions. \\
In all cases we find a rapidly spinning central object (period $\sim$ 1 ms) with masses (depending on the initial spins) between 2.5 and 3.1 M$_{\odot}$. This object might be stable on the simulation time scale, but we suspect it to collapse to a black hole within milliseconds after the merger. 
Our central objects are surrounded by thick disks containing (depending on the initial spins) $\sim$ 0.1 to 0.3 M$_{\odot}$. In addition we find long extended tails for the corotating models. These expand explosively when the neutron star matter gets decompressed by tidal torques in the LS-EOS cases when the adiabatic exponent rises and  a few times $10^{49}$ erg are released due to the recombination of nucleons into heavy nuclei. For the polytropic EOS ($\Gamma= 2.6$) the tails remain thin and well-defined. For the other initial spins both the central object and the disk are embedded in a low density cloud of decompressed neutron star matter. In all cases (apart from the one where the stars spin against the orbit) almost baryon free funnels form above the poles of the central object. In these funnels large gradients of radiation pressure could be built up that would be able to accelerate indrifting matter into two jets pointing away from the poles. These funnels would also be an ideal place for relativistic fireballs to form. However, a mass as small as $10^{-5}$ M$_{\odot}$ within such a fireball would be enough to prevent a GRB from forming. The present resolution is too low to draw conclusions on this point. We find SPH-smoothed temperatures of up to 50 MeV. These are found in macroscopic vortex structures  that form along the contact surface of both stars and which we suspect to originate from Kelvin-Helmholtz instabilities.\\
The main new result is the amount of mass that is ejected into space. We find that, dependent on the initial spins, between $4\cdot10^{-3}$ and $4\cdot10^{-2}$ M$_{\odot}$ become unbound for the realistic equation of state of Lattimer and Swesty. This result is strongly dependent on the EOS, a stiff polytrope ejects only around one half of this material. In the test case of a soft polytrope ($\Gamma= 2.0$) we cannot resolve any mass loss, which indicates a strong sensitivity on the stiffness of the EOS.
The material gets ejected at very high densities ranging from $10^{12}$ to $10^{14}{\rm g cm^{-3}}$. The bulk of matter gets ejected with Ye below 0.05 with small contaminations of the neutron star crust ($Y_{e} \approx 0.3$). Such low $Y_{e}$, low entropy matter is prone to form r-process nuclei. However, the results concerning $Y_{e}$ are biased by our simple neutrino treatment. Using recent rates for neutron star mergers we find that  $\sim 3 \cdot 10^{-3}$ to $0.3  \; {\rm M}_{\odot}$ of r-process material would have to be ejected to explain the observed abundances exclusively by coalescing neutron stars. Thus our numbers for the amount of ejecta look promising and
if, as suggested, large parts of this matter consist of r-process nuclei, neutron star mergers could account for all the observed r-process material in the Galaxy.

\acknowledgements
It is a pleasure to thank P. H\"oflich and H.-T. Janka for useful discussions
and M. Guidry for carefully reading the manuscript.
This work was  supported by the Swiss National 
                        Science Foundation under grants No. 20-47252.96 and 
                        2000-53798.98 and in part by the National Science 
                        Foundation under 
			Grant No. PHY 94-07194 (S.R., M.L., F.-K.T.), M.B.D. 
			acknowledges the support of the 
                        Royal Society through a University
                        Research Fellowship, part of the work by W.B. has been 
                        supported by the Swiss National
                        Science Foundation,
			T.P. was supported by the US-Israel BSF grant 95-328 .

\begin{appendix}


\section{The Particle Setup}
In the initial setup of the SPH-particles we adopt an intermediate approach between the extremes of total randomness and the well defined order of a lattice. In a first step we create  a large set of spherical shells, containing initially randomly distributed particles which have been relaxed according to some Coulomb-type force law. The so constructed shells are then joined in a concentric way, and the particles are assigned masses that, together with the smoothing length, reproduce the desired density profile. This configuration is finally relaxed with the SPH-code.\\To construct a single shell of a given particle number  we distribute the particles randomly on the surface of a unit sphere. We assume some repulsive, dissipative Coulomb-type force law:
\begin{eqnarray}
\vec{f}_{i}  =  \sum_{j \neq i} \frac{\vec{r}_{i j}}{|\vec{r}_{i j}|^{3}} - \alpha \vec{v}_{i},
\end{eqnarray}
where 
$\alpha$ is some dissipation parameter, $\vec{r}_{i j}$ is the difference vector of particle $i$ and $j$, $\vec{r}_{i j}= \vec{r}_{i} - \vec{r}_{j}$, $ \vec{r}_{i} $ and  $\vec{v}_{i}$ are particle position and velocity vectors.
In the following relaxation process only the force component tangential to the spherical surface  was used in order to constrain the movement to a constant radius. After each integration time step the particle positions were projected back onto the surface.\\
To be able to vary the particle number density with the radius we introduce  a parameter $0< \epsilon < 1$. The distance to the new shell $\Delta r_{n+1}$ is then calculated from the previous one $\Delta r_{n}$ via
\begin{equation}
\Delta r_{n+1}= \epsilon \Delta r_{n}.
\end{equation}
Denoting the distance from the center to the first shell by $\Delta r_{1} \equiv \delta$ and requiring that
\begin{equation}
R_{{\rm star}} \stackrel{!}{=} \sum_{i=1}^{N_{{\rm shell}}+1} \Delta r_{i},
\end{equation}
$R_{{\rm star}}$ is the radius of our star,
we find for a given number of shells $N_{{\rm shell}}$ and given $\epsilon \ne 1$:
\begin{eqnarray}
\delta  =  R_{{\rm star}} \frac{1-\epsilon}{ 1-\epsilon^{N_{{\rm shell}}+1}}.
\end{eqnarray}
Now the question arises of how to choose the particle number in a given shell. We have as a constraint that typical particle separations within shell $n$ should be the same as typical distances between shells $\Delta \bar{r}_{n}= \frac{\Delta r_{n}+\Delta r_{n+1}}{2}$. We consider now a sphere of radius $R$ with $N_{n}$ SPH-particles distributed on its surface. If we calculate the typical distance of two particles within one shell and equate it to $\Delta \bar{r}_{n}$ we find for the particle number in shell n
\begin{equation}	
N_{n}= \frac{2}{1- \sqrt{1- \frac{\Delta \bar{r}_{n}^{2}}{4 R^{2}}} } \label{npart}. 
\end{equation}
In the case of $N_{n} \gg 1$ we have $\frac{\Delta \bar{r}_{n}}{R} \ll 1$ and can therefore expand the square root in the denominator:
\begin{equation}	
N_{n}\approx 16 \frac{R^{2}}{\Delta \bar{r}_{n}^{2}}
 \end{equation}
The same result is recovered by assigning to each particle the surface of a circle and setting this equal to $\frac{4 \pi R^{2}}{N_{n}}$.\\
To find the optimal relative orientation of the shells 
we go back to the idea of the Coulomb-type forces.\\
The innermost shell is kept fixed. Then the torque exerted from shell one to shell two  is calculated and shell two is rotated (according to the corresponding Eulerian angles) until the torque vanishes. 
Fixing the second shell at the new position, the torque on shell three  from the inner shells  is calculated and so on.\\
The profiles $\rho^{p}(r)$, $Y_{e}^{p}(r)$ and $u^{p}(r)$ define our initial neutron star model. They are calculated using a Newtonian 1D stellar structure code and a self-consistent table of $\rho$, $Y_{e}$ and $u$ generated using the LS-EOS. For the calculation of $Y_{e}$, we use (\ref{mue}) (at $T= 4 \cdot 10^{9}$ K) and the $\beta$-equilibrium condition for the chemical potentials:
\begin{equation}
\mu_{e}= \hat{\mu} + Q - \mu_{\bar{\nu}_{e}},
\end{equation}
where $Q$ is the mass difference of neutron and proton times $c^{2}$ and the difference in the chemical potentials of neutron and proton 
(without rest masses), $\hat{\mu}$, is provided by the LS-EOS. Since we want to construct a cold neutron star where no neutrinos are present, we assume their chemical potentials to vanish,
\begin{equation}
\mu_{\bar{\nu}_{e}}= 0.
\end{equation}
The $Y_{e}$-profiles are for numerical reasons restricted to values $\gid$ 0.05.\\
After values of $Y_{e}$ and $u$ have been assigned to the particles according to the profiles, the masses still have to be distributed. In principle 
\begin{equation}
\vec{\rho}= W \vec{m},
\end{equation}
has to be solved, where the components of the vectors are the values at the particle positions 
(density and mass) and $W$ is the kernel matrix.\\
In order to enforce sphericity, we assign one single value for all the masses $m_{i}$ and one for all the smoothing lengths $h_{i}$ to all the particles in a given shell.
We  prescribe an allowed range of neighbours for the mean neighbour number in each shell (the neighbours are found by traversing the tree of our SPH-code) and adjust the smoothing lengths correspondingly (typical neighbour numbers are $\approx 65$). Starting with the masses from the density self-contribution of each particle
\begin{equation}
m_{i}= \pi h_{i}^{3} \; \rho^{p}(\vec{r}_{i}),
\end{equation}
where the spherical spline kernel (see Benz (1990)) has been used, the masses  are iterated until
\begin{equation}
\max_{j} \left( \frac{|\bar{\rho_{j}} - \rho^{p}(r_{j})|}{\rho^{p}(r_{j})} \right) < \epsilon_{\rho},
\end{equation}
where $\bar{\rho_{j}}$ is the mean value of the density in shell $j$ ($\epsilon_{\rho}$ was typically a few times $10^{-3}$). \\
This particle setup was then finally relaxed with the SPH-code where a velocity dependent additional force term was applied:
\begin{equation}
	\vec{f}_{i,{\rm hydro}} \rightarrow \vec{f}_{i,{\rm hydro}} - \gamma \vec{v}_{i}.
\end{equation}
The parameter $\gamma$ gives a characteristic damping time scale
$\tau = \gamma^{-1}.$
To obtain an efficient convergence towards the equilibrium solution it is important to choose $\gamma$ to be slightly over-critical, i.e.
$\tau \le \tau_{{\rm osc}}$,
where the typical oscillation period $\tau_{{\rm osc}}$ is approximately given by the sound crossing time of the neutron star
$\tau_{{\rm osc}} \approx \tau_{s} = (G \bar{\rho})^{-\frac{1}{2}} \approx 0.3 \; {\rm ms}$. Some properties of the relaxed star are shown in Fig. \ref{finsetup}.

\section{The Equation of State}
For all our calculations the default set of nuclear parameters (see Lattimer $\&$ Swesty (1991)) with $K= 180$ MeV is used.\\
Most of the scheme that we are going to describe here has been developed for version 2.6 of the LS-EOS. We encountered several regions (especially for low $Y_{e}$ and densities where  large amounts of the heavy nuclei are present) where the Newton-Raphson iteration to solve the set of equilibrium equations did not converge for the default set of guess values (nucleon number density inside nuclei $n_{i}$, nucleon degeneracy parameters $\eta_{n}$ and $\eta_{p}$) due to very small convergence radii.\\
To find accurate guess values $g_{i}$ in the above mentioned critical regions we use the fact that the $g_{i}$ are slowly varying functions of $Y_{e}$.
We chose  a fixed, uncritical value of $Y_{e}$, $Y^{{\rm ref}}_{e}= 0.25$, for which reference surfaces of the guess values $G_{i,Y_{e}^{{\rm ref}}}$ over the $n-T$-plane, $n$ is the nucleon number density, $T$ the temperature, were calculated

\begin{equation}
G_{i,Y_{e}^{{\rm ref}}}: (n,T) \rightarrow g_{i}.
\end{equation}
To patch the surface over the critical region for some given $Y_{e}$, the corresponding rectangular piece in the reference surface was cut out and matched to the hole in the surface of the desired $Y_{e}$. Let

\begin{equation}
(n_{1},n_{2}) \times (T_{1},T_{2})
\end{equation}
be the region where the code  does not converge for a given $Y_{e}$. We then calculate multipliers $m_{i,j}$ for each guess value $g_{i}$ 

\begin{equation}
m_{i,j}(Y_{e})= \frac{ g_{i,j}(Y_{e})}{g_{i,j}(Y_{e}^{{\rm ref}})},
\end{equation}
with $j$ labeling the edge point, in such a way that the edge points in the reference surface coincide with those for the critical $Y_{e}$ when multiplied with the corresponding $m_{i,j}$:

\begin{equation}
g_{i,j}(Y_{e})= m_{i,j}(Y_{e}) \; \; g_{i,j}(Y_{e}^{{\rm ref}}).
\end{equation}
Then a multiplier $m_{i}$ for the points within the critical region is calculated by means of linear interpolation:

\begin{eqnarray}
m_{i}(n,T,Y_{e})&=&(1-x)(1-y) \; m_{i,1}(Y_{e})\; +\; x \;(1-y) \; m_{i,2}(Y_{e})\nonumber \\ 
&& + \; x y \; m_{i,3}(Y_{e})\; + \; (1-x) y \; m_{i,4}(Y_{e}) ,
\end{eqnarray}
where $x= (n-n_{1})(n_{2}-n_{1})^{-1}$ and $y= (T-T_{1})(T_{2}-T_{1})^{-1} $ and the edge points are numbered starting with ($n_{1},T_{1}$) in a counter-clockwise sense.
The guess value variables in the critical region are then approximated by
\begin{equation}
g_{i}(n,T,Y_{e})= m_{i}(n,T,Y_{e}) \cdot  g_{i}(n,T,Y_{e}^{{\rm ref}}).
\end{equation}
Using this scheme accurate tables for the guess values have been calculated. Using these tables the code converges rapidly and safely everywhere.\\
In our hydrodynamic calculations we use a tabular form of the LS-EOS (version 2.7), where the above described scheme has been used. Our table contains 153 entries in $\log(\rho)$, 121 entries in $\log(u)$, and 25 entries in $Y_{e}$. For our calculations we need to tabulate the pressure, the temperature and the difference in the nucleon chemical potentials, $\hat{\mu}$. The abundances are difficult to tabulate since they may vary enormously from grid point to grid point. Therefore, every time that abundances are needed (e.g. to calculate the total nuclear binding energy of the system), the original LS-EOS is used.\\
Since we use the specific internal energy as independent variable, we performed a bisection iteration to find the corresponding temperatures. If non-monotonic values in $T$ were encountered, the iteration used $u$ values that were averaged over neighboring values. Note, however, that apart from that, no other manipulation of the EOS-data has been performed (Ruffert et al. (1996) used smoothed EOS-tables).\\
The interpolation in the table is a delicate topic in itself since thermodynamic consistency for the values between  the tabulated grid points is not guaranteed. An inconsistent interpolation, i.e. an interpolation that does not fulfill the constraints posed by the Maxwell relations of thermodynamics, will lead to an artificial buildup of entropy, or temperature. Swesty  (1996) has proposed an interpolation scheme insuring thermodynamic consistency as well as the continuity of the derivatives of the thermodynamic functions. However, this approach applied to our 3D-input parameter space would require 216 terms and the evaluation of fourth order derivatives for each table call. Clearly, this procedure is - at least at present - computationally prohibitive. Considering the other approximations in our model, we think that we can justify just a linear interpolation in our table.

\section{Neutrino Emission Rates}
Starting from the electron capture cross section (see Tubbs and Schramm (1975)), which reads with our approximations

\begin{equation}
\sigma_{EC}= \frac{ 1  +  3  \; \alpha^{2}} {8} \; \sigma_{0} \; \frac{E_{\nu_{e}}^{2}}{(m_{e} c^{2})^{2}},
\end{equation}
where $\alpha \approx 1.25$, ${E_{\nu_{e}}}$ is the energy of the emitted neutrino, 

\begin{equation}
\sigma_{0}= \frac{4  \; G_{F}^{2}  \; m_{e}^{2}  \; \hbar^{2}}{\pi  \; c^{2}}
\end{equation}
and $G_{F}$ the Fermi constant,
the number of electron captures per time and volume, $R_{EC}$, is given by ($k_{B}=1$)

\begin{equation}
R_{EC} =  \eta_{pn} \; \frac{\pi}{h^{3} c^{2}} \;  \frac{1  +  3   \alpha^{2}}{(m_{e} c^{2})^{2}} \; \sigma_{0}  \; T^{5} \; F_{4}[\eta_{e}].
\end{equation}
The corresponding energy loss in neutrinos per time and volume is 

\begin{equation}
Q_{EC}= \eta_{pn} \; \frac{\pi}{h^{3} c^{2}} \frac{1  +  3  \; \alpha^{2}}{(m_{e} c^{2})^{2}} \; \sigma_{0} \;  T^{6} \; F_{5}[\eta_{e}].
\end {equation}
Here $\eta_{pn}= \rho N_{A} (Y_{n}-Y_{p})/(e^{\frac{\mu_{n}-\mu_{p}}{T}}-1)$, with the $Y_{i}$ denoting neutron and proton  abundances, $\mu_{i}$ the corresponding chemical potentials (without rest mass) and $N_{A}$ Avogadro's constant. The factor $\eta_{pn}$, derived  by Bruenn (1985) under the assumption that $4$-momentum transfer between leptons and nucleons is negligible (``elastic approximation''), takes into account effects resulting from the phase space  restrictions of the final state nucleons.
$F_{5}$ is the usual Fermi integral given by

\begin{equation}
F_{n}(z)= \int_{0}^{\infty} \frac{x^{n}}{e^{x-z}+1} {\rm d}x.
\end{equation}
$\eta_{e}$ is the electron degeneracy parameter, $\eta_{e}=\frac{\mu_{e}}{T}$.
The electron chemical potential is calculated according to 
\begin{eqnarray}
\mu_{e} =  kT (\rho N_{A})^{\frac{1}{3}} \; \{  [ \sqrt{b^{2} + a^{3}}+ b]^{\frac{1}{3}} 
 -   [ \sqrt{b^{2} + a^{3}} - b]^{\frac{1}{3}}   \} \label{mue},
\end{eqnarray}
where  $a \equiv  (\pi^{2}/3 - \beta^{2}/2) (\rho N_{A})^{-2/3}$, $b \equiv \frac{3 h^{3}}{16 \pi} (\frac {c}{T})^{3} Y_{e}$, $\beta \equiv m_{e} c^{2}T^{-1}$ have been introduced (see Baron (1985)).\\
Assuming $\mu_{e^{+}}= -\mu_{e}$ the corresponding formulae for the positron captures read 

\begin{equation}
R_{PC} = \eta_{np} \; \frac{\pi} { h^{3} c^{2}} \frac{1 + 3  \; \alpha^{2}}{(m_{e} c^{2})^{2}} \; \sigma_{0} \; T^{5} \; F_{4}[- \eta_{e}]
\end{equation}

and 

\begin{equation}
Q_{PC} = \eta_{np} \; \frac{\pi} { h^{3}c^{2} } \frac{1 + 3  \; \alpha^{2}}{(m_{e} c^{2})^{2}} \; \sigma_{0} \; T^{6} \; F_{5}[- \eta_{e}],
\end{equation}
where $\eta_{np}$ is found from $\eta_{pn}$ by interchanging the indices. For simplicity we  assume here that matter consists only of protons and neutrons:

\begin{equation}
Y_{p}= Y_{e}, \quad Y_{n}=1-Y_{e}.
\end{equation}
The rate of change in the electron fraction is  given by

\begin{equation}
\dot{Y_{e}}= \frac{R_{PC}}{n_{n}} Y_{n} -  \frac{R_{EC}}{n_{p}} Y_{p}.
\end {equation}
\end{appendix}
\clearpage

\newpage
\bibliographystyle{apj}
\bibliography{literat}

\clearpage

\begin{center}
\begin{figure}[h]
\psfig{file=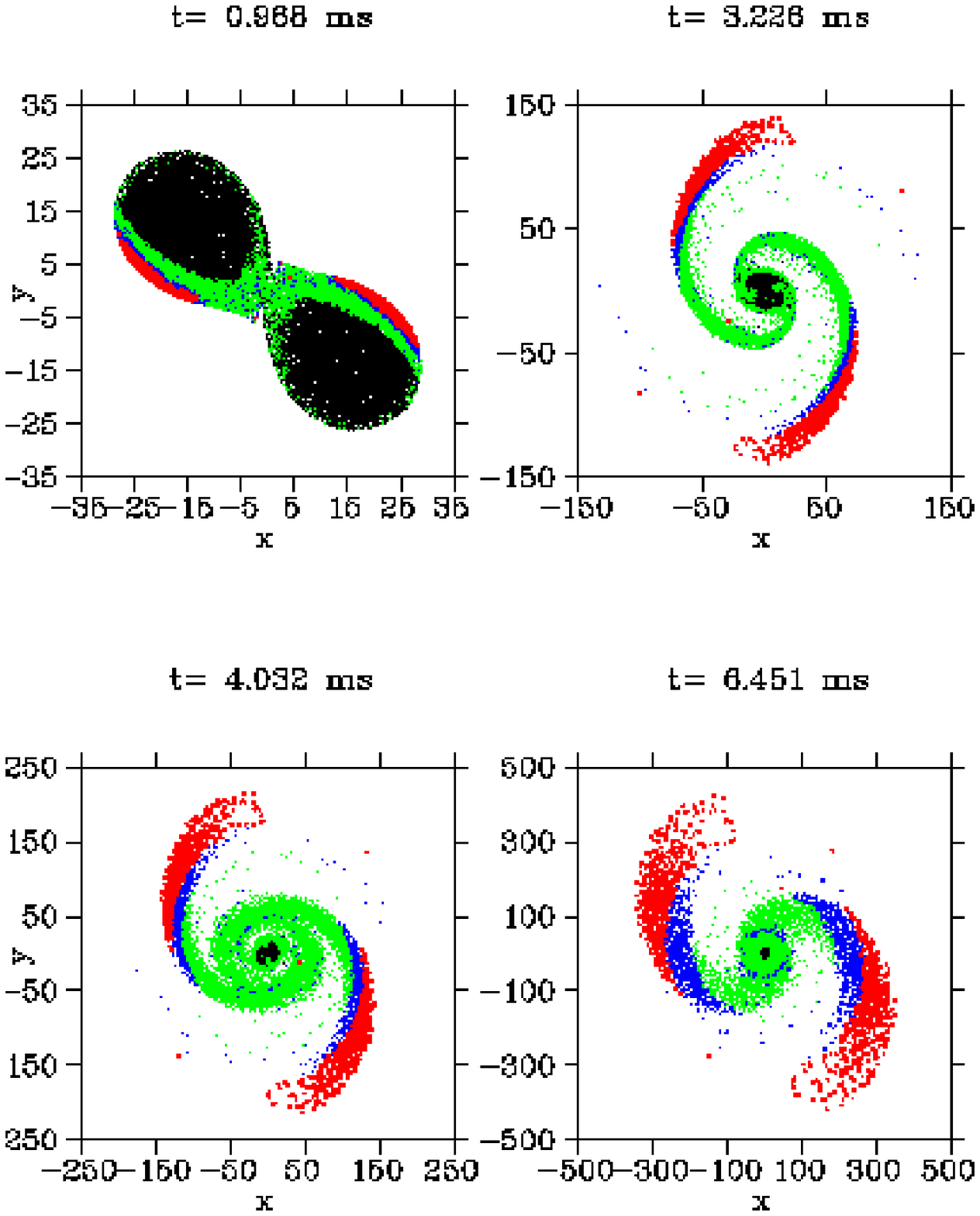,width=14cm,angle=0}
\caption{\label{morphA1} Morphology of run A (representative for corotation).}
\end{figure} \clearpage

\begin{figure}[h]
\psfig{file=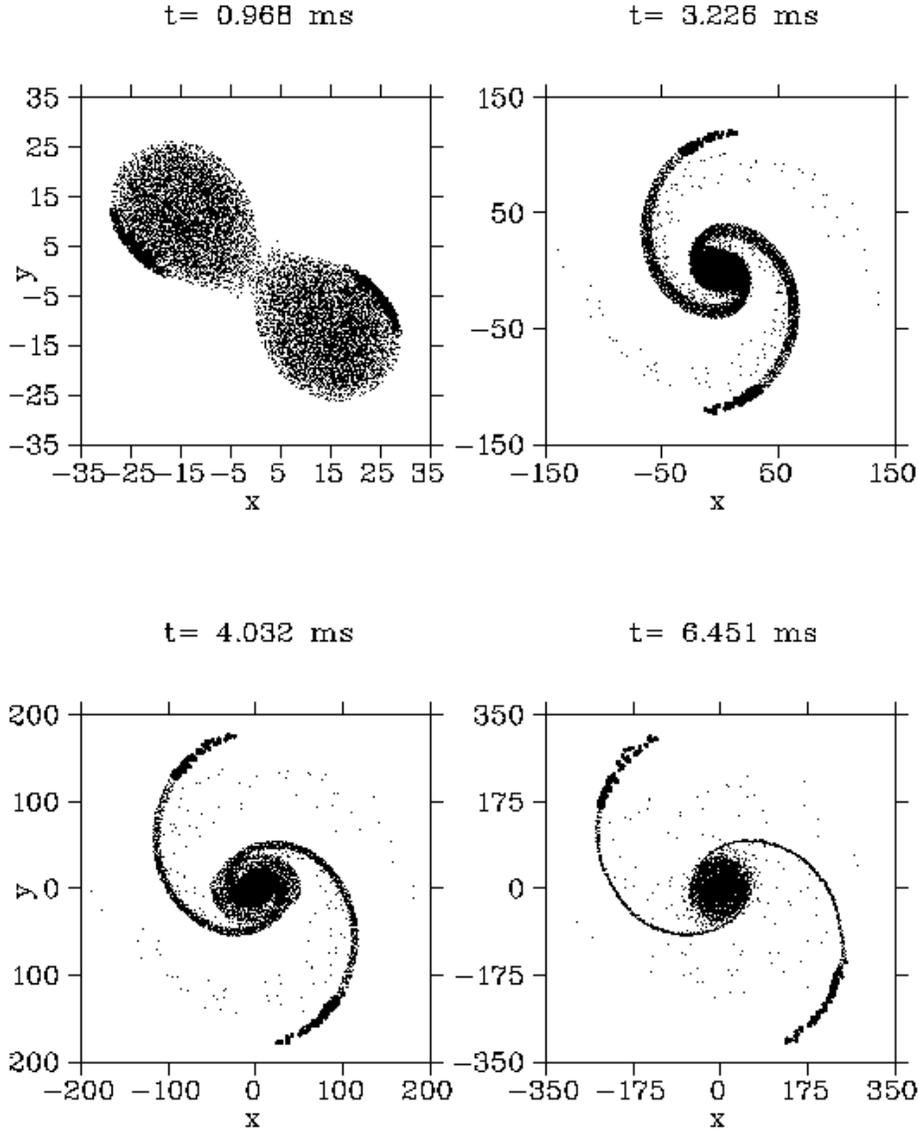,width=14cm,angle=0}
\caption{\label{morphC1} Morphology of run C (corotation, polytropic EOS with $\Gamma= 2.6$); the
 large dots denote the positions of the escaping particles.}
\end{figure} \clearpage

\begin{figure}[h]
\psfig{file=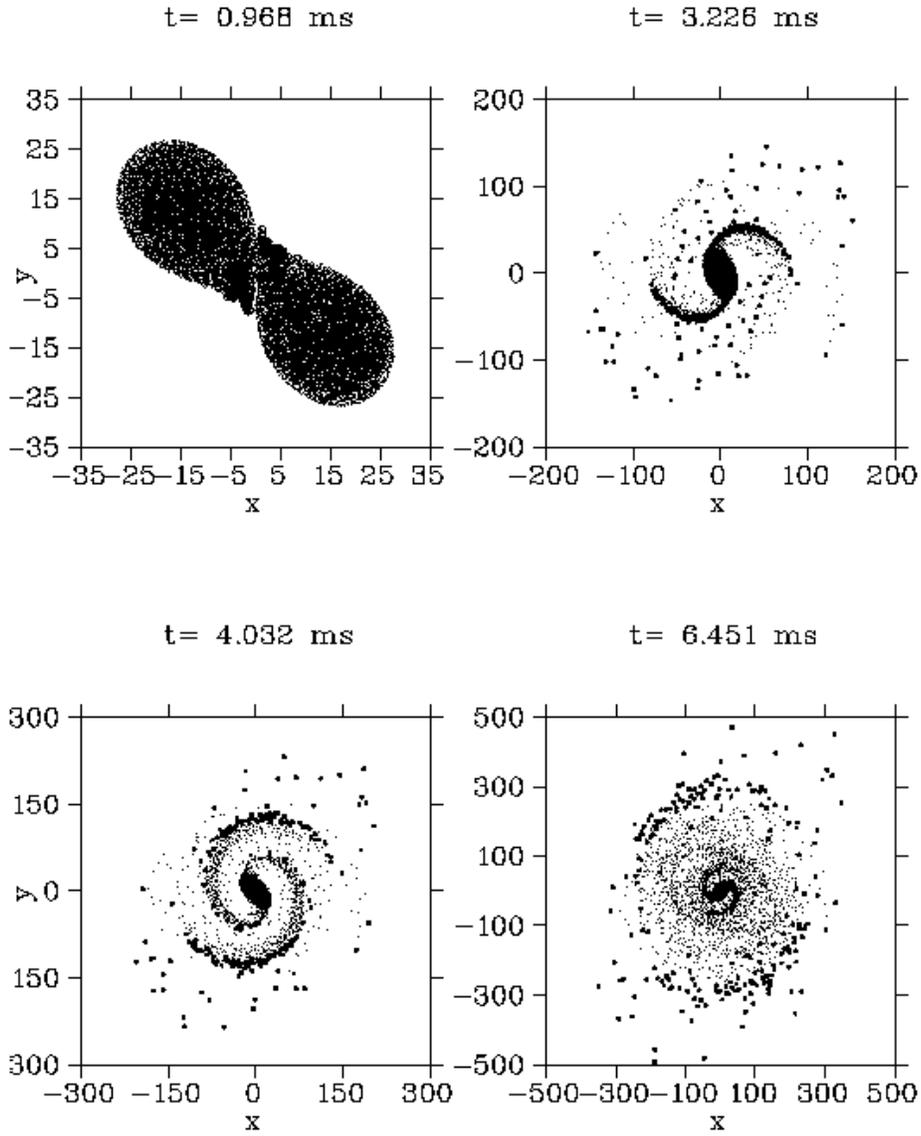,width=14cm,angle=0}
\caption{\label{morphI1} Morphology of run I (no initial spin).}
\end{figure} \clearpage

\begin{figure}[h]
\psfig{file=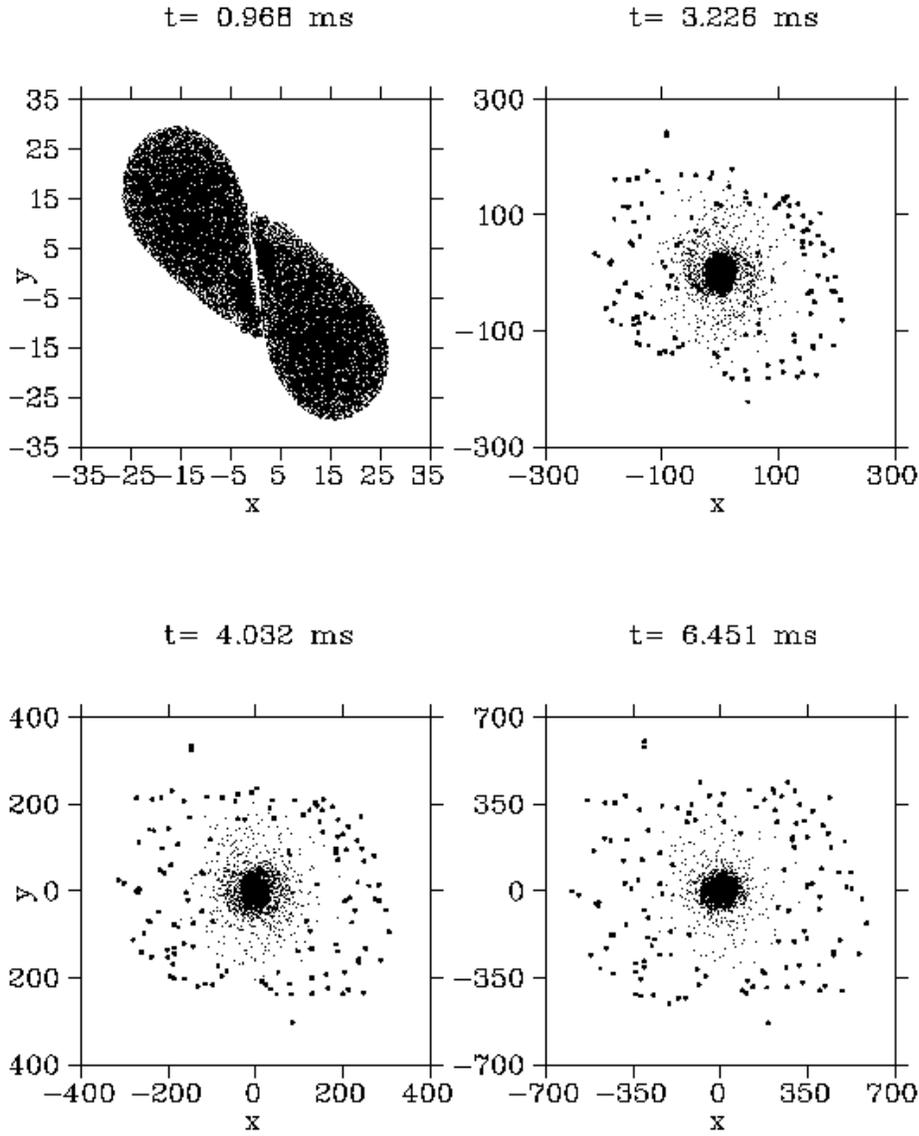,width=14cm,angle=0}
\caption{\label{morphJ1} Morphology of run J (spins against the orbital motion). }
\end{figure} \clearpage

\begin{figure}[h]
\psfig{file=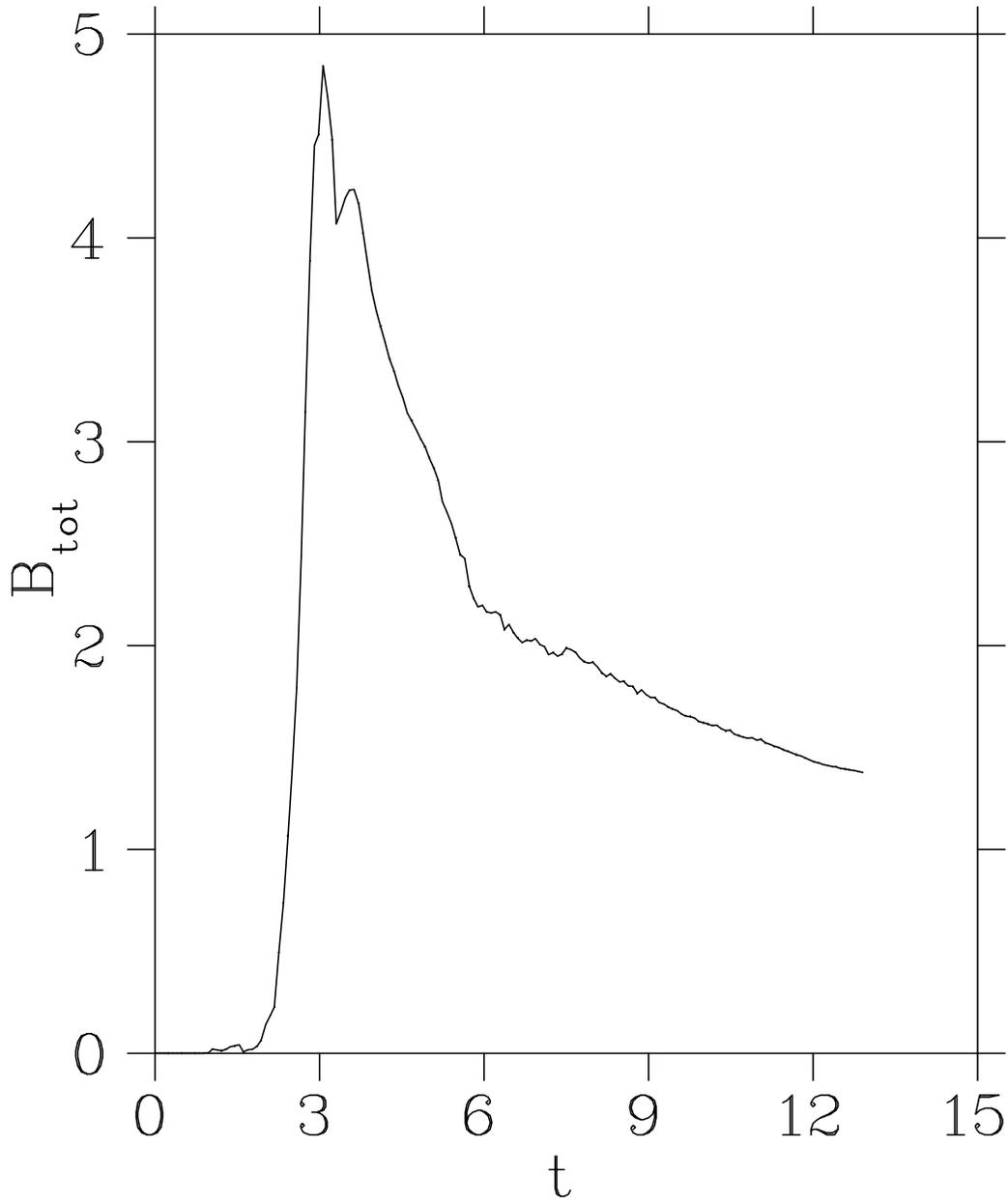,width=14cm,angle=0}
\caption{\label{bnuctot} Total amount of nuclear energy (in units of $10^{50}$ erg) in the system (time is given in ms).}
\end{figure} \clearpage

\begin{figure}[h]
\psfig{file=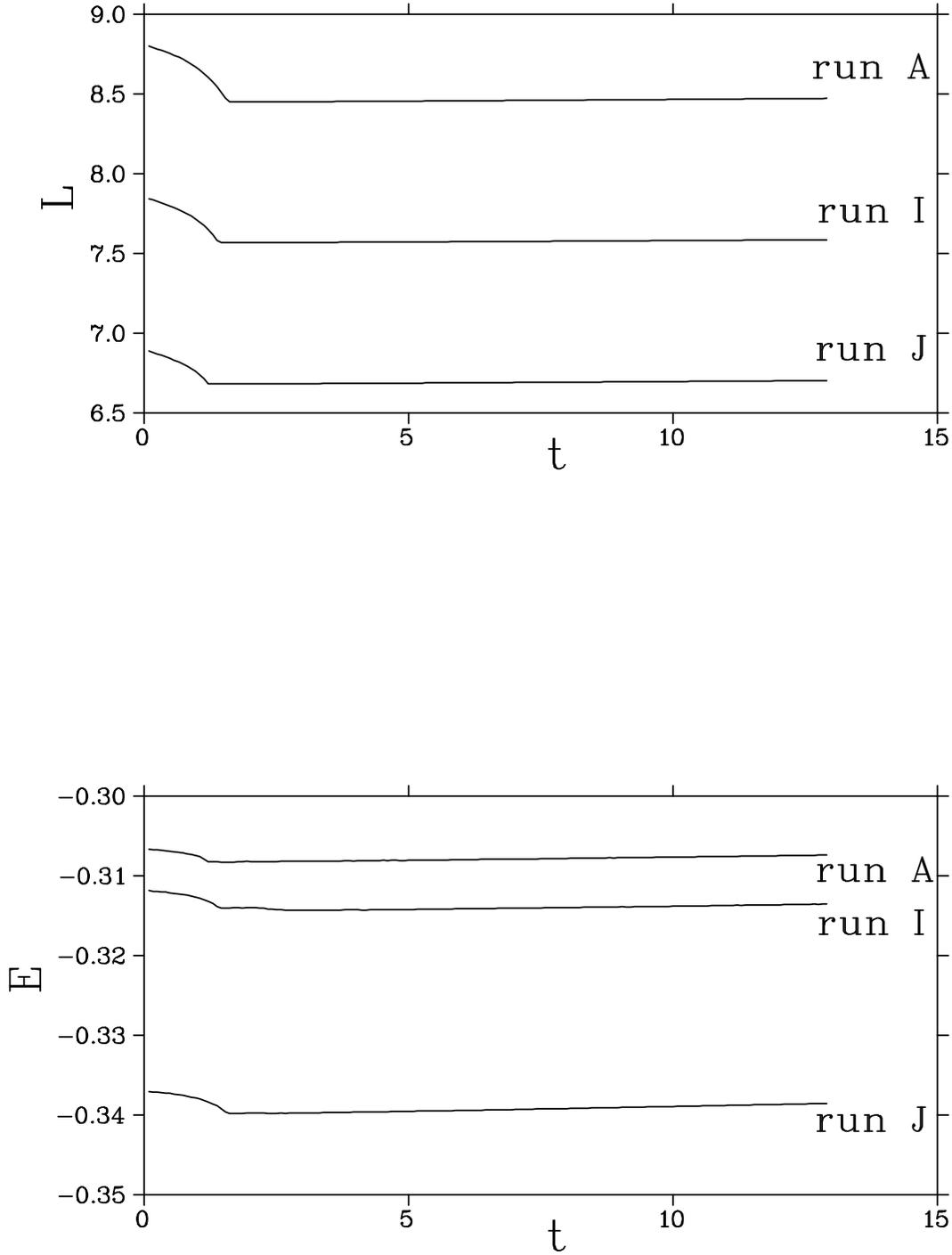,width=14cm,angle=0}
\caption{\label{conservation} The panels show the evolution of total angular momentum L and total energy E of runs A, I and J. L and E are conserved to approximately $3\cdot10^{-3}$ (the decrease in the beginning corresponds to the phase where angular momentum as well as energy are lost through the emission of gravitational waves).}
\end{figure} \clearpage

\begin{figure}[h]
\psfig{file=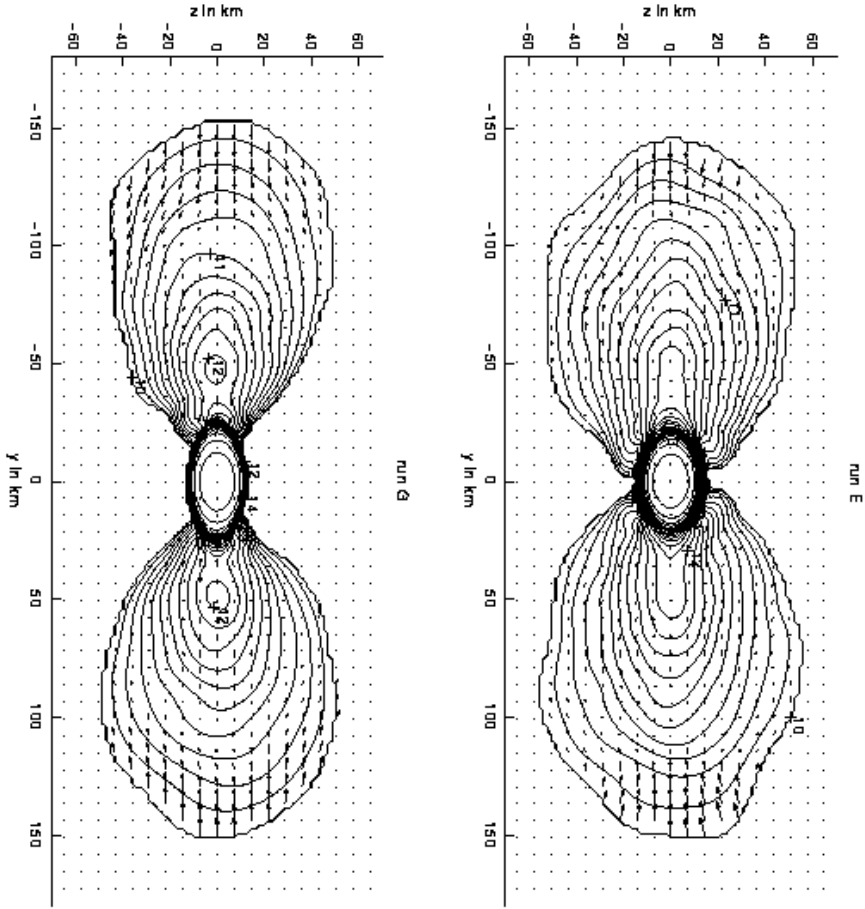,width=14cm,angle=0}
\caption{\label{yzbw1}Cut through the y-z-plane of the last dumps of the runs E (corotation, no neutrinos) and G (corotation, neutrinos in free-streaming limit); the labels in all density contour plots refer to $\log(\rho_{[{\rm g cm}^{-3}]})$.}
\end{figure} \clearpage

\begin{figure}[h]
\psfig{file=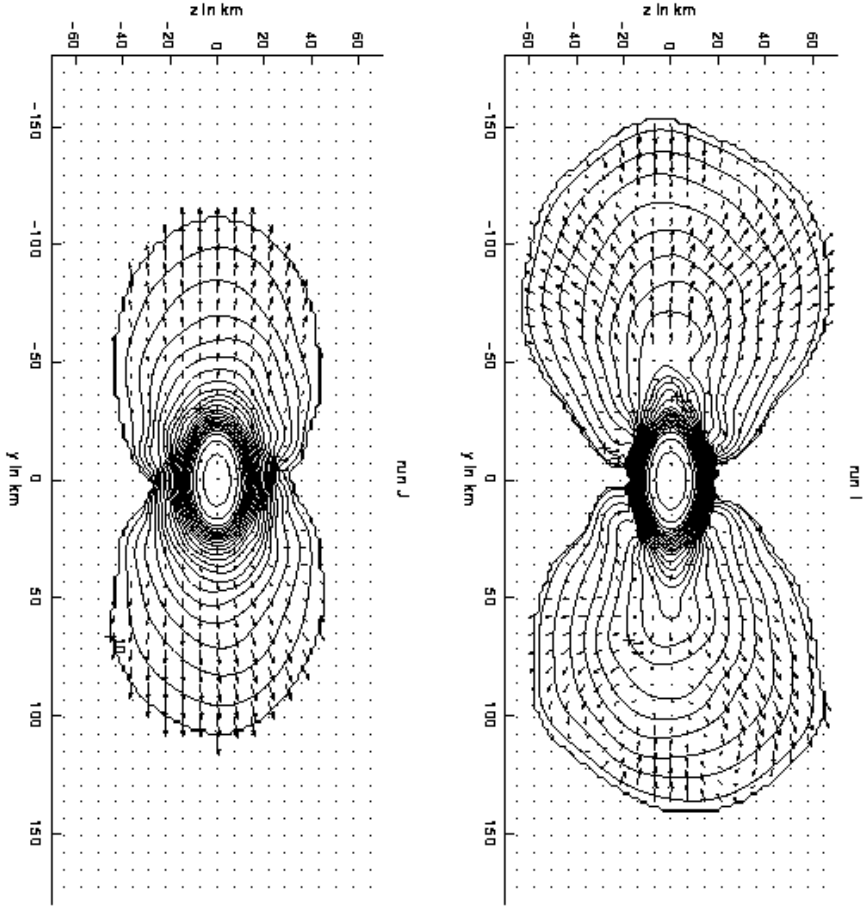,width=14cm,angle=0}
\caption{\label{yzbw2} Last dumps of the runs I (no initial spins) and J (spins against orbit).}
\end{figure} \clearpage

\begin{figure}[h]
\psfig{file=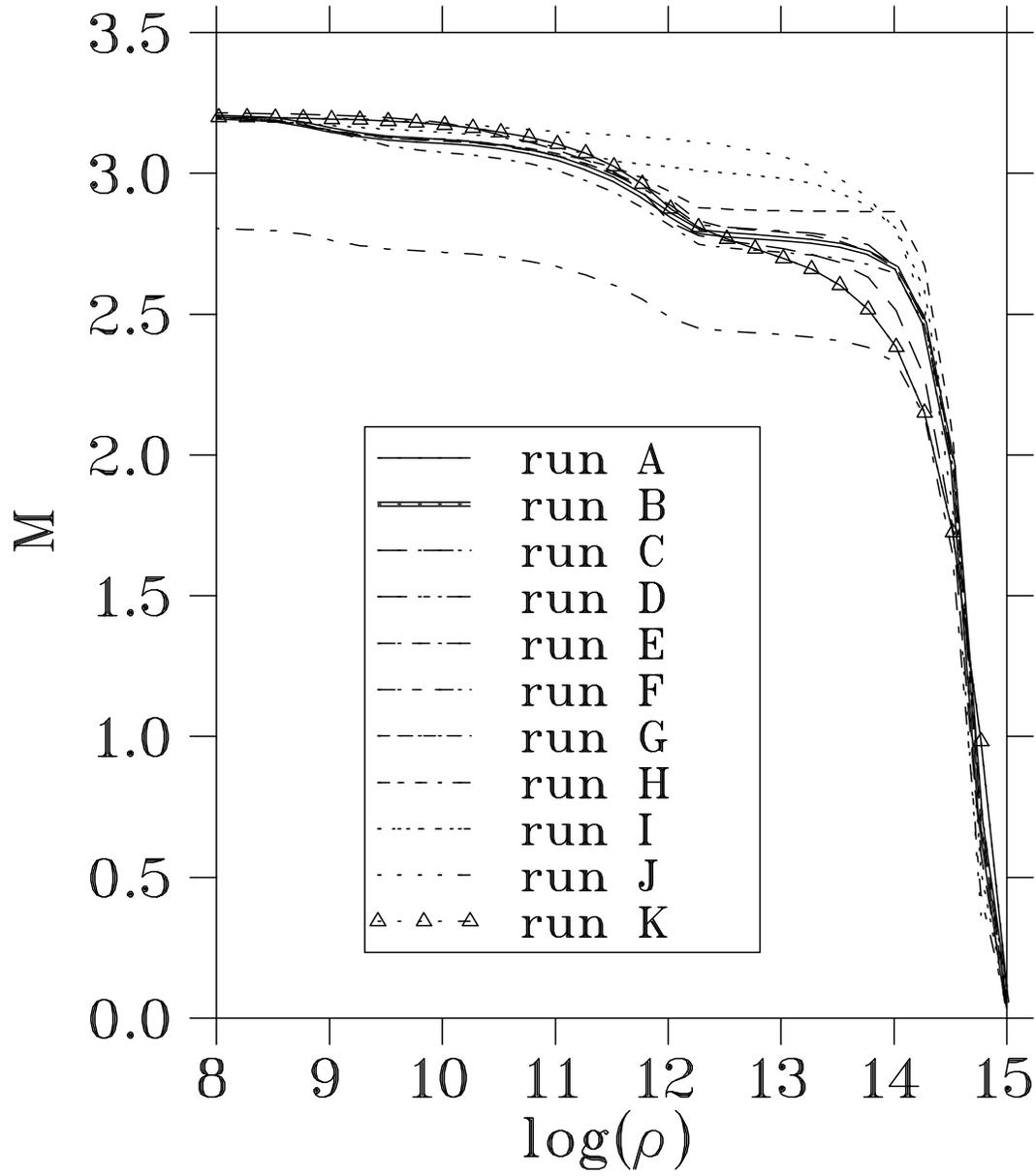,width=14cm,angle=0}
\caption{\label{mvsrho} The figure shows the mass that has a larger density than log$(\rho)$ for all the runs (last dump).}
\end{figure} \clearpage

\begin{figure}[h]
\begin{center}
\psfig{file=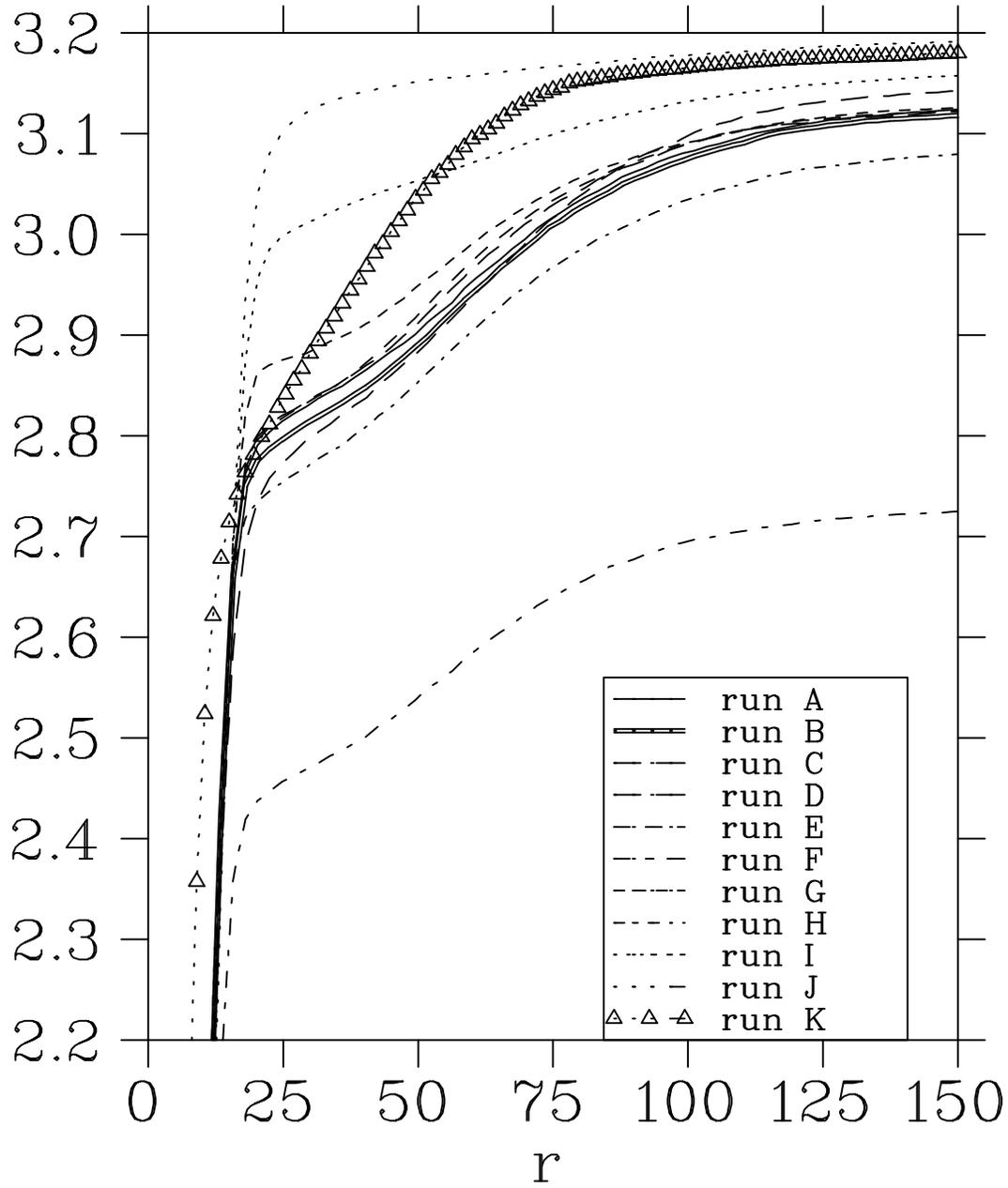,width=14cm,angle=0}
\caption{\label{mvsr} Distribution of mass with cylindrical radius (last dump). Masses are given in solar units, the radius in km.}
\end{center}
\end{figure} \clearpage

\begin{figure}[h]
\psfig{file=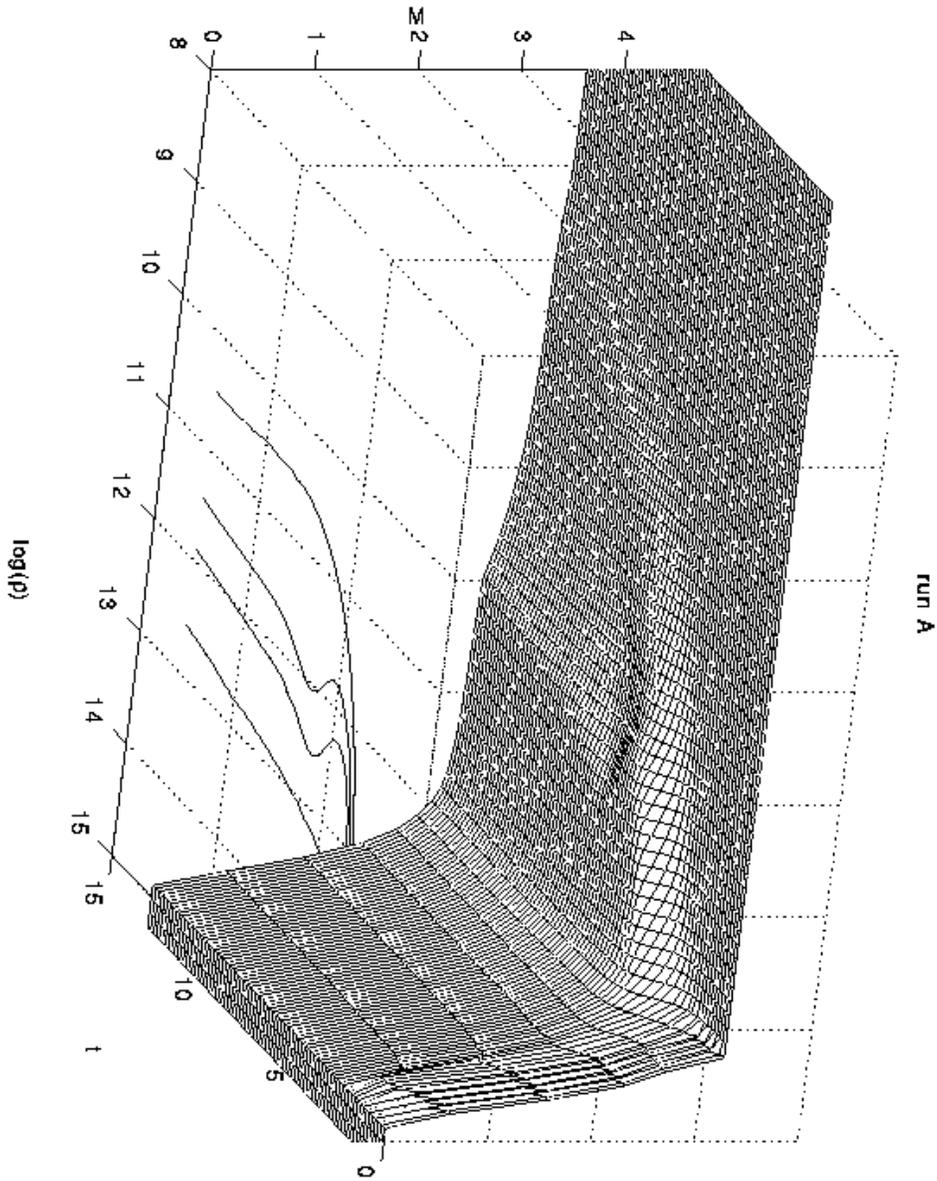,width=14cm,angle=0}
\caption{\label{mvsrhovst} Time evolution (run A, corotation) of the mass that has a larger density than log$(\rho)$. The contour lines in the t-${\rm log}(\rho)$-plane correspond to 2.8, 2.9, 3.0 and 3.1 M$_{\odot}$. Time is given in 
ms, $\rho$ in g cm$^{-3}$.}
\end{figure} \clearpage

\begin{figure}[h]
\psfig{file=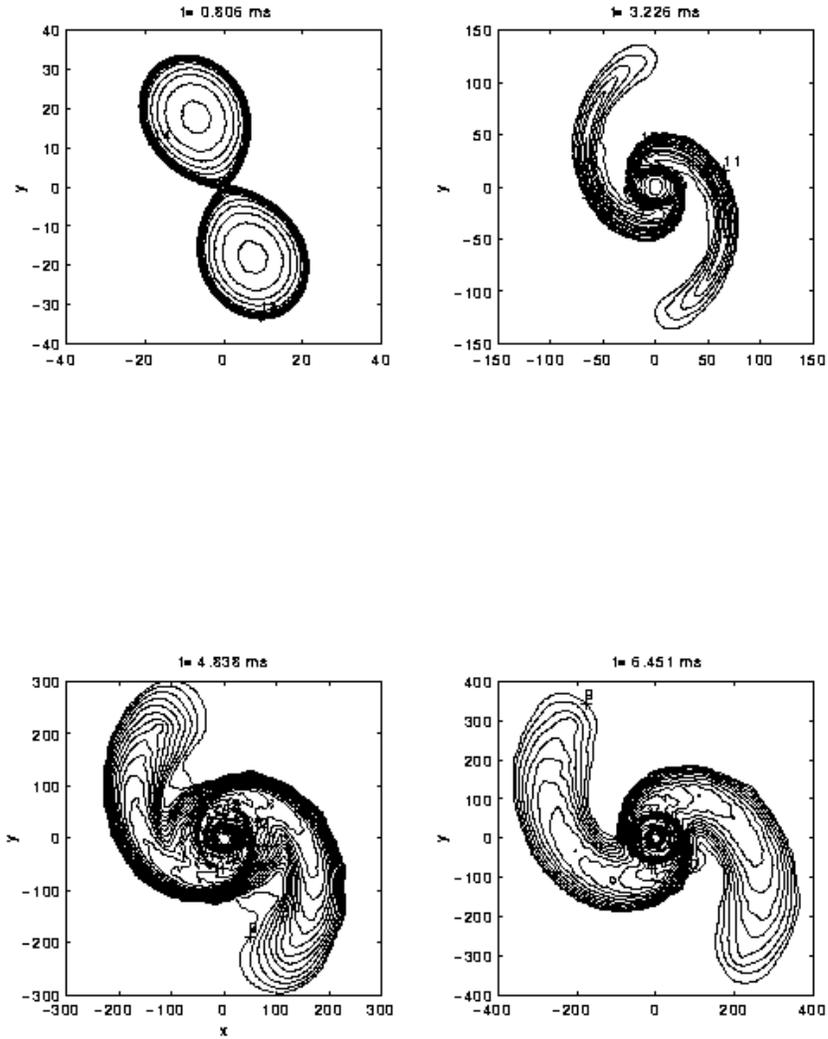,width=14cm,angle=0}
\caption{\label{xyEbw} Density contours  of run E (corotation), lengths are given in km.}
\end{figure} \clearpage

\begin{figure}[h]
\psfig{file=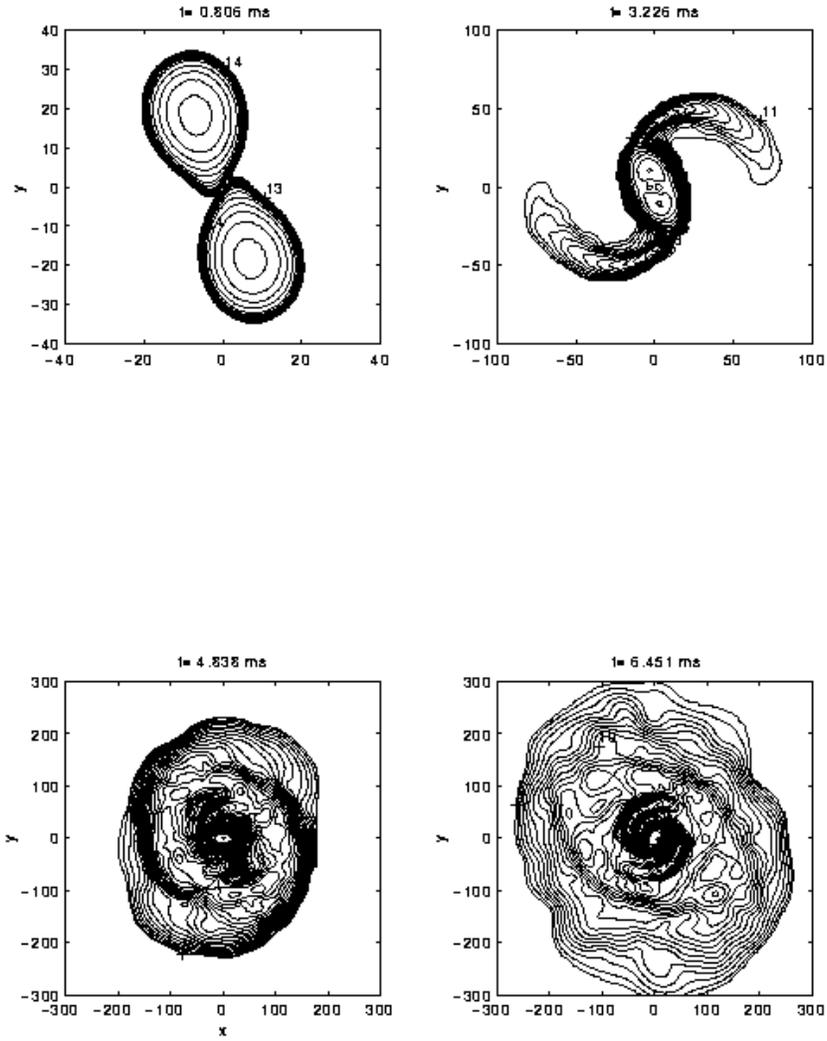,width=14cm,angle=0}
\caption{\label{xyIbw} Density contours  of run I (no initial spins), lengths are given in km.}
\end{figure} \clearpage

\begin{figure}[h]
\psfig{file=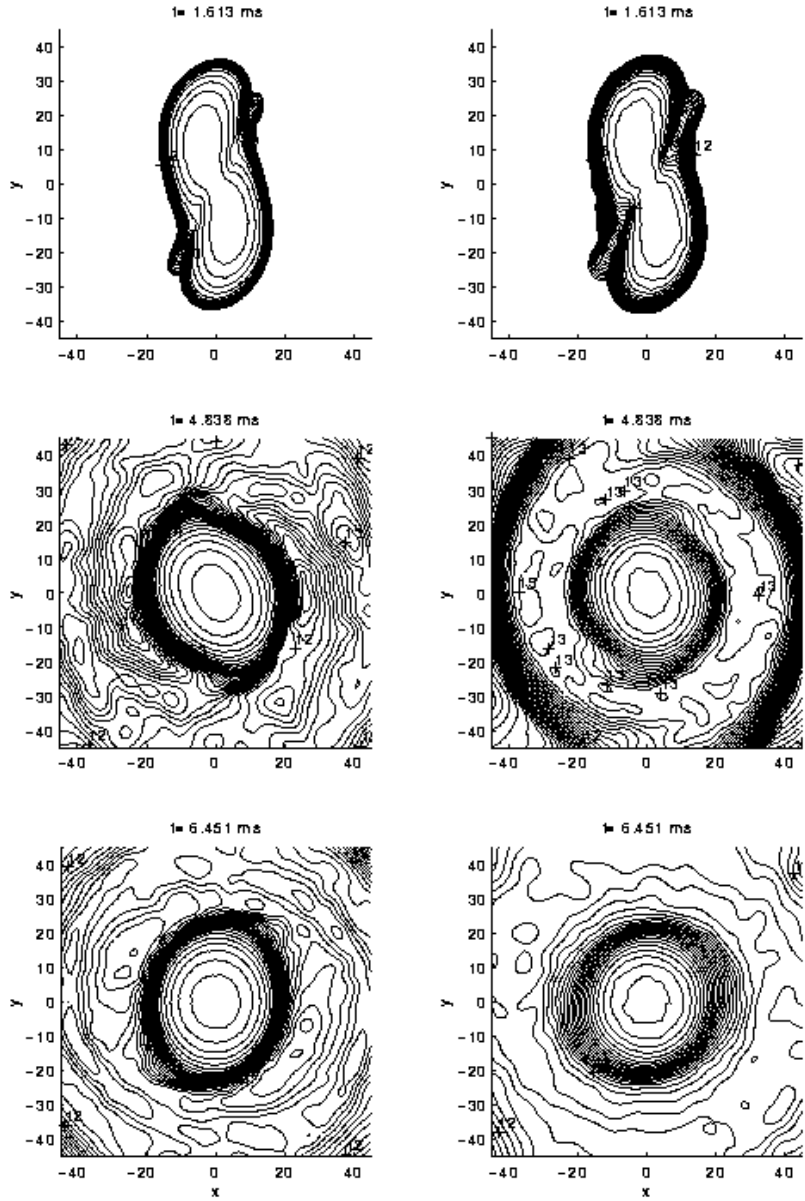,width=14cm,angle=0}
\caption{\label{xyECbw} Density contours of the runs E (LS-EOS, left column) and C (polytrope with $\Gamma= 2.6$, right column), lengths are given in km, the contour labels refer to $\log(\rho_{[{\rm g cm}^{-3}]})$.}
\end{figure} \clearpage

\begin{figure}[h]
\psfig{file=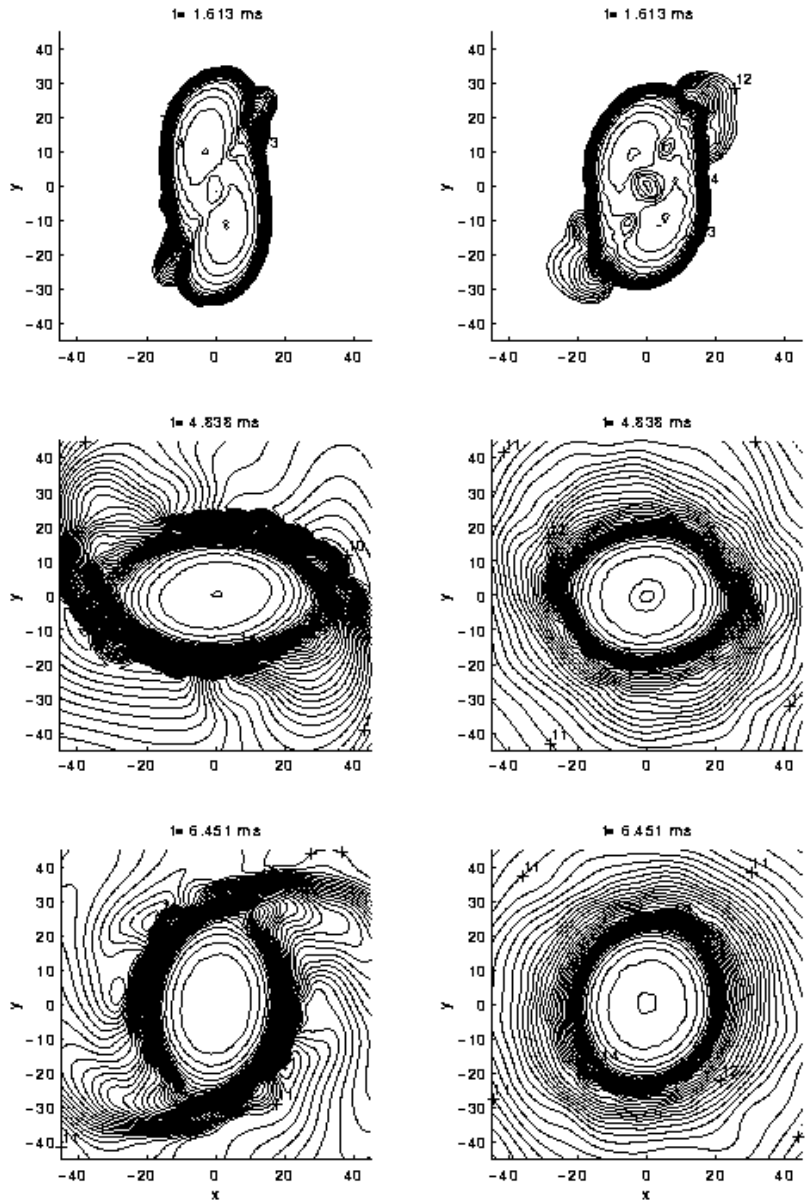,width=14cm,angle=0}
\caption{\label{xyIJbw} Cut through the x-y-plane of the runs I (no spins, left column) and J (spins against orbit, right column), lengths are given in km.}
\end{figure} \clearpage

\begin{figure}[h]
\psfig{file=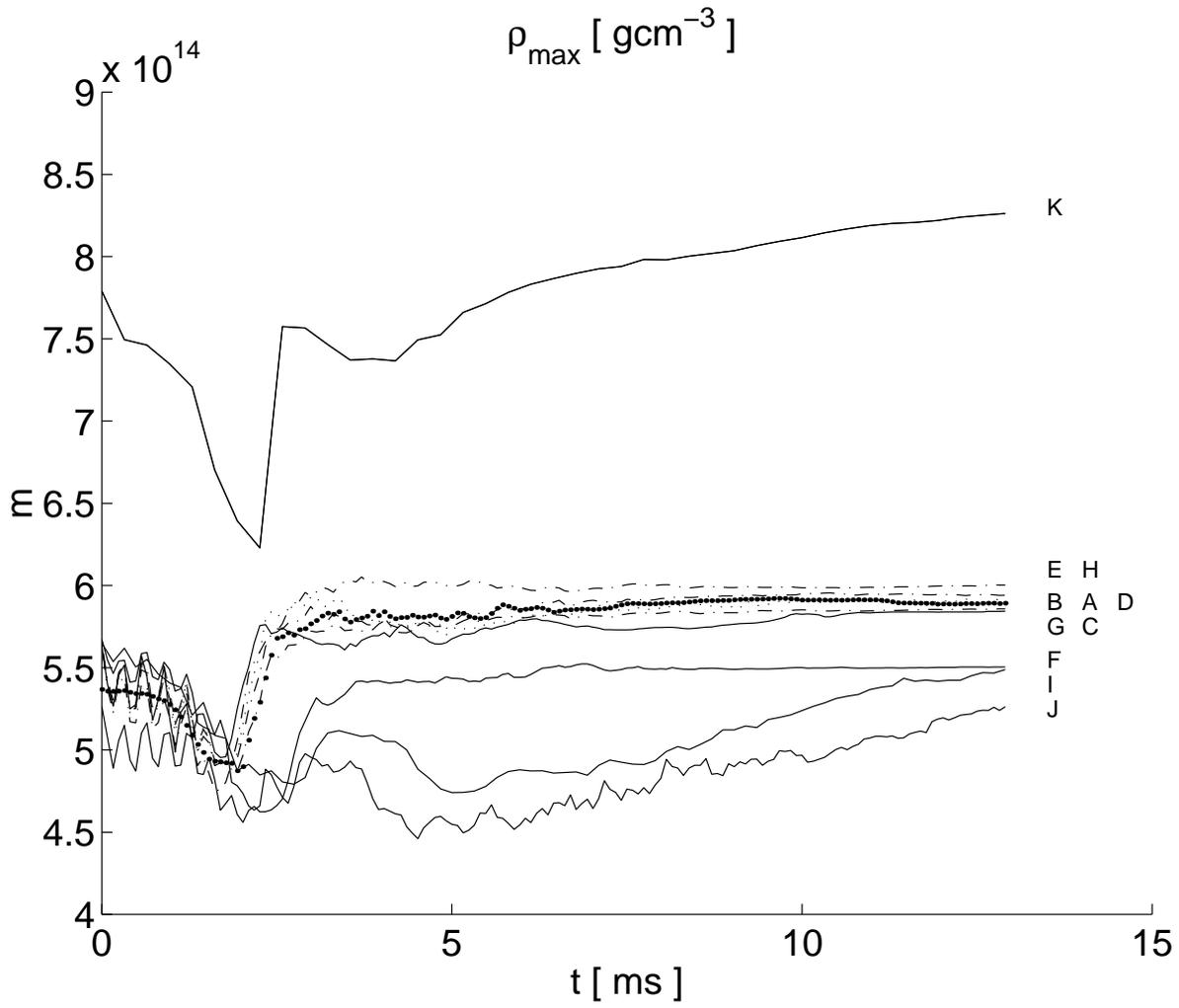,width=16cm,angle=0}
\caption{\label{rhomult} Maximum densities (in units of $ {\rm g cm}^{-3}$) obtained in our runs.}
\end{figure} \clearpage

\begin{figure}[h]
\psfig{file=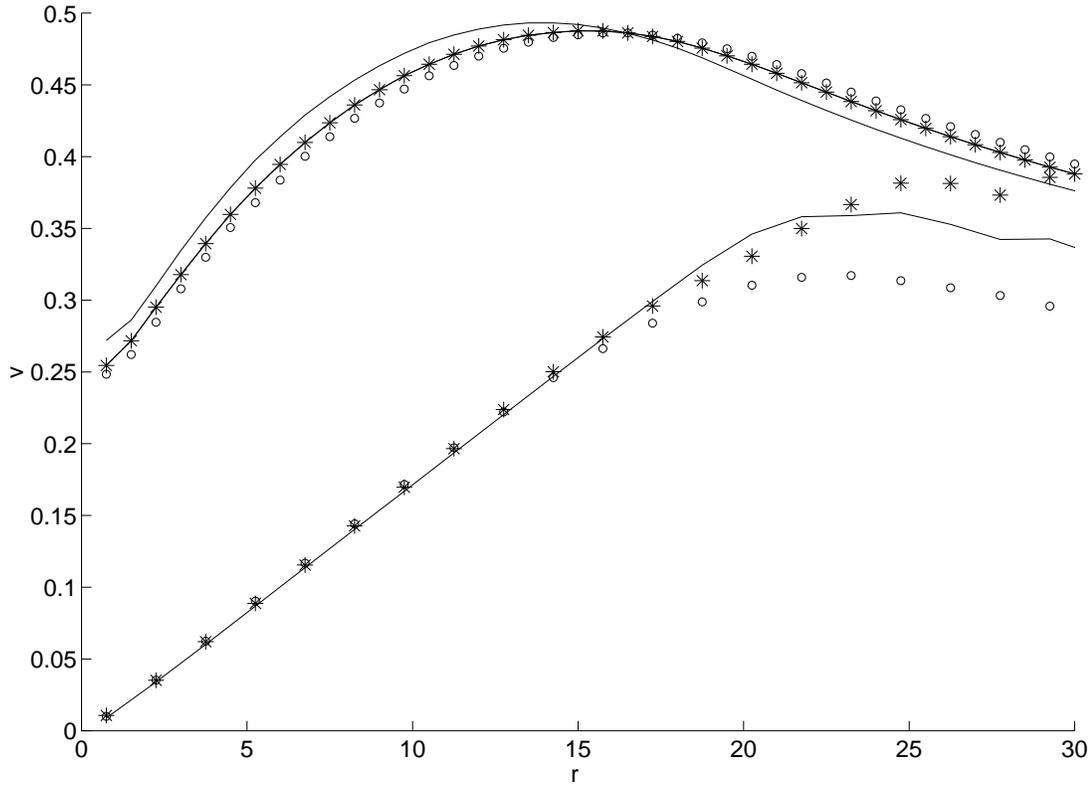,width=16cm,angle=90}
\caption{\label{velcomp} The upper three curves correspond to the Kepler-velocities ($c=1$), the lower ones are mean tangential particle velocities versus cylindrical radius. The solid line corresponds to corotation (run E), the circles to run I (no spins) and the asterisks to run J (spins against orbit).}
\end{figure} \clearpage

\begin{figure}[h]
\psfig{file=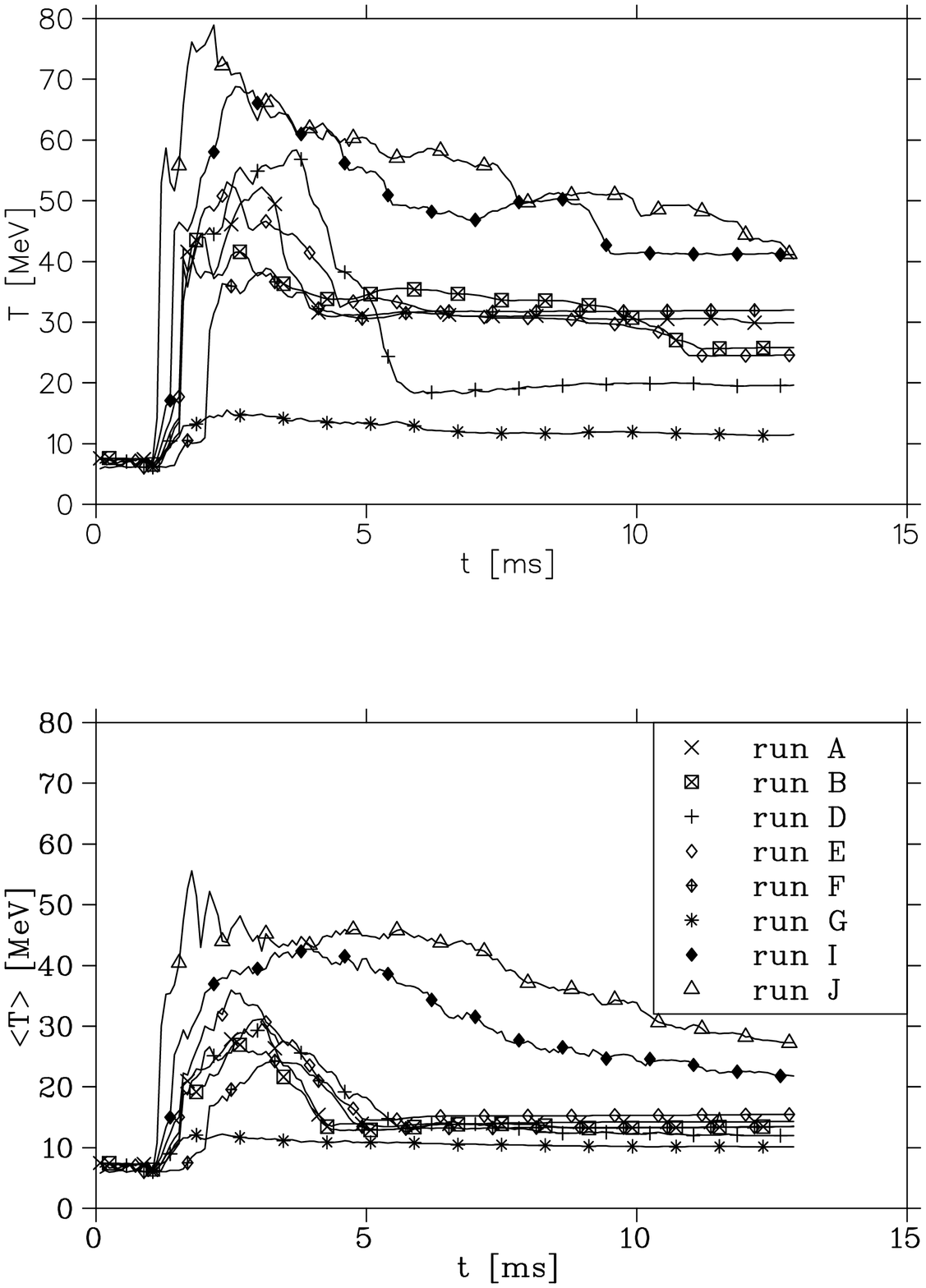,width=14cm,angle=0}
\caption{\label{Tmax} Shown are the maximum temperatures of the different runs. The upper panel shows the maximum particle temperatures, the lower one the maximum of the SPH-smoothed values. The legend refers to both panels, temperatures are given in units of MeV.}
\end{figure} \clearpage

\begin{figure}[h]
\psfig{file=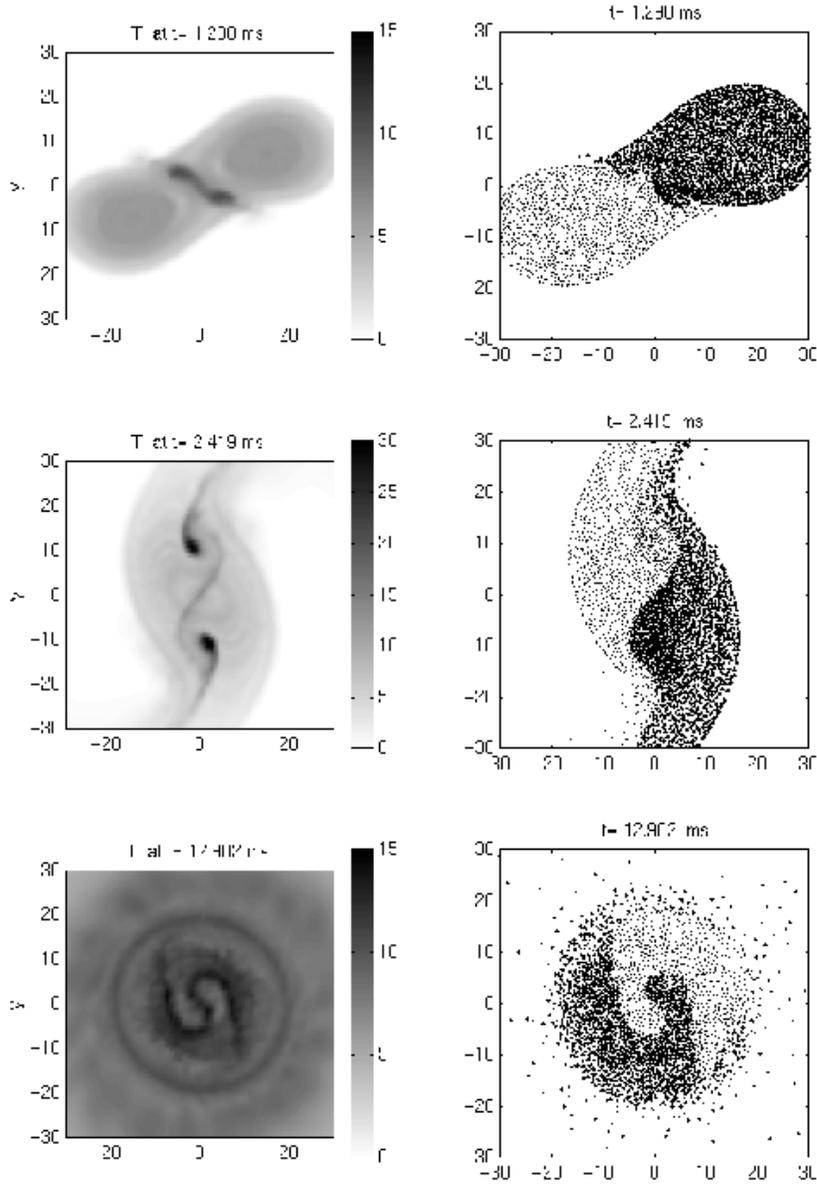,width=14cm,angle=0}
\caption{\label{TmixE} The left column shows the SPH-smoothed temperatures (in MeV) of run E (corotation). The right column shows the origin of all particles from either of the stars (crosses, star 1; dots, star 2).}
\end{figure} \clearpage

\begin{figure}[h]
\psfig{file=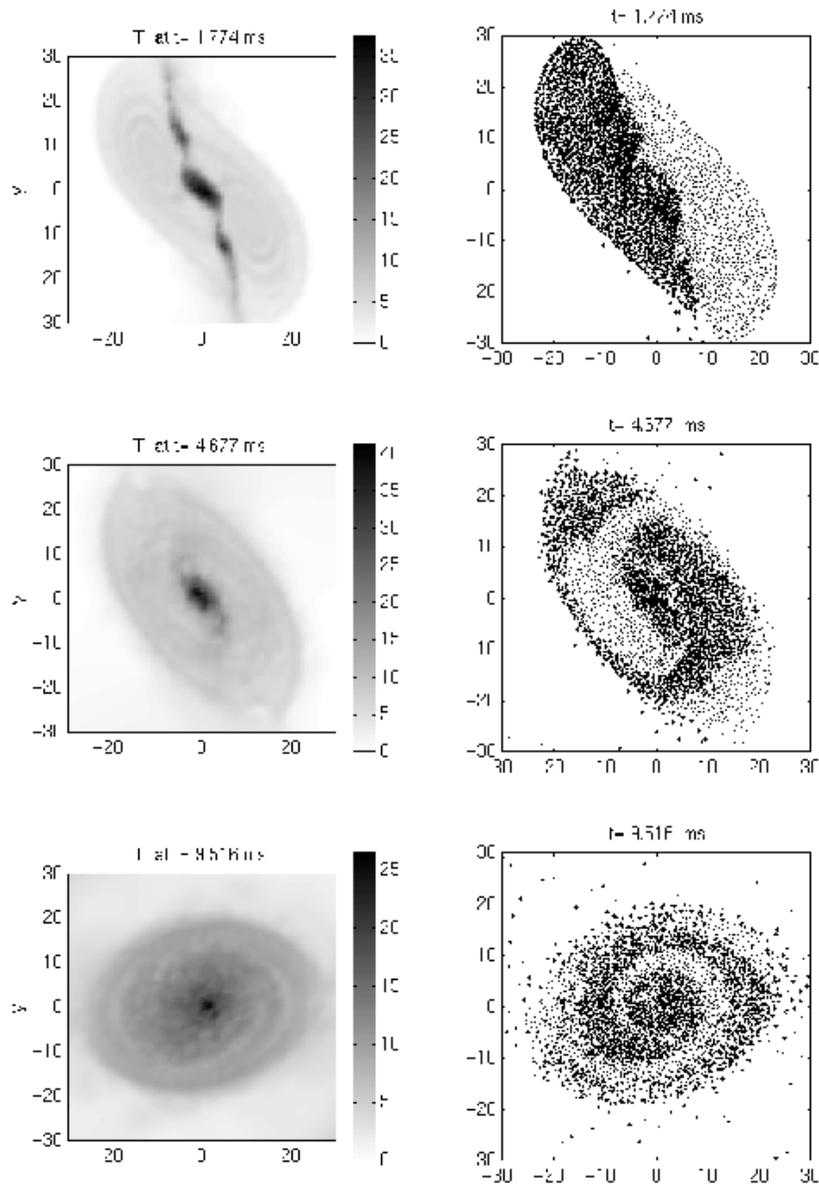,width=14cm,angle=0}
\caption{\label{TmixI} SPH-smoothed temperatures (in MeV) of run I (no initial spins). The particles from each star are plotted with different symbols.}
\end{figure} \clearpage

\begin{figure}[h]
\psfig{file=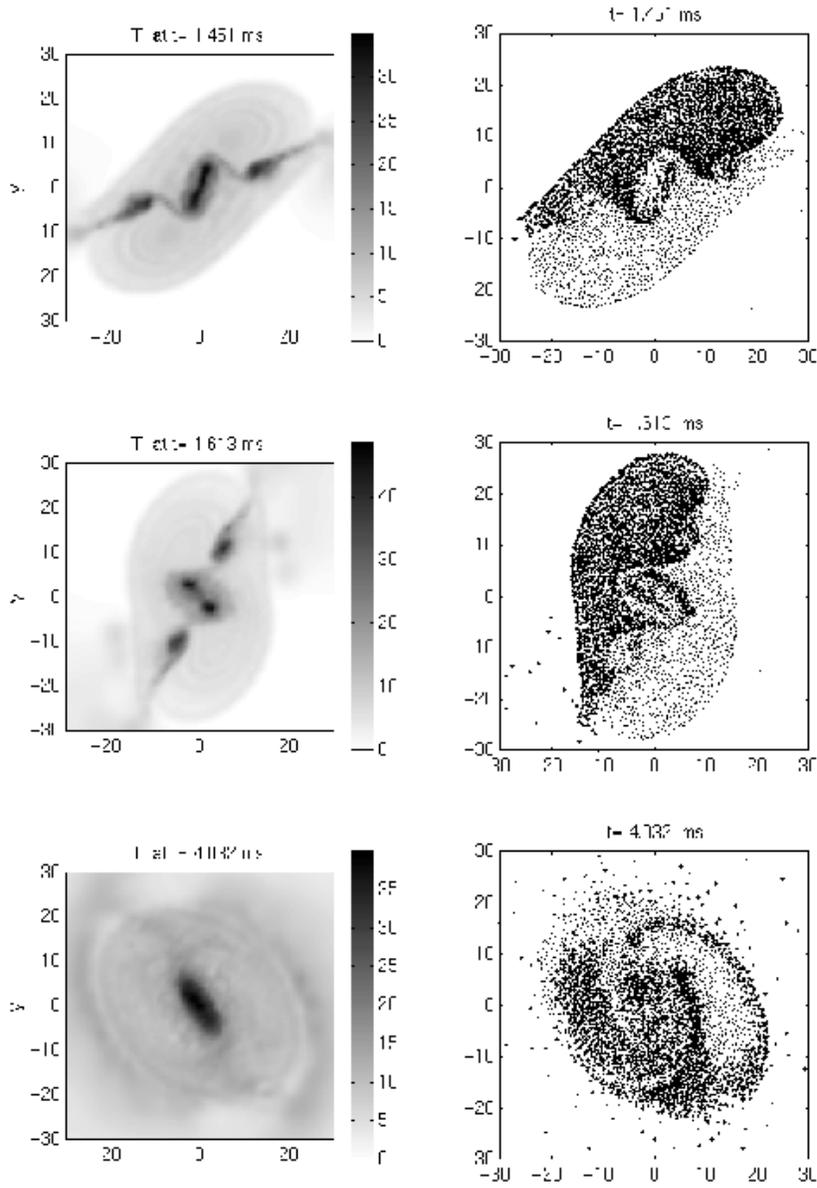,width=14cm,angle=0}
\caption{\label{TmixJ} Same as Fig. \ref{TmixI}, but for run J (spins against orbital angular momentum).}
\end{figure} \clearpage

\begin{figure}[h]
\psfig{file=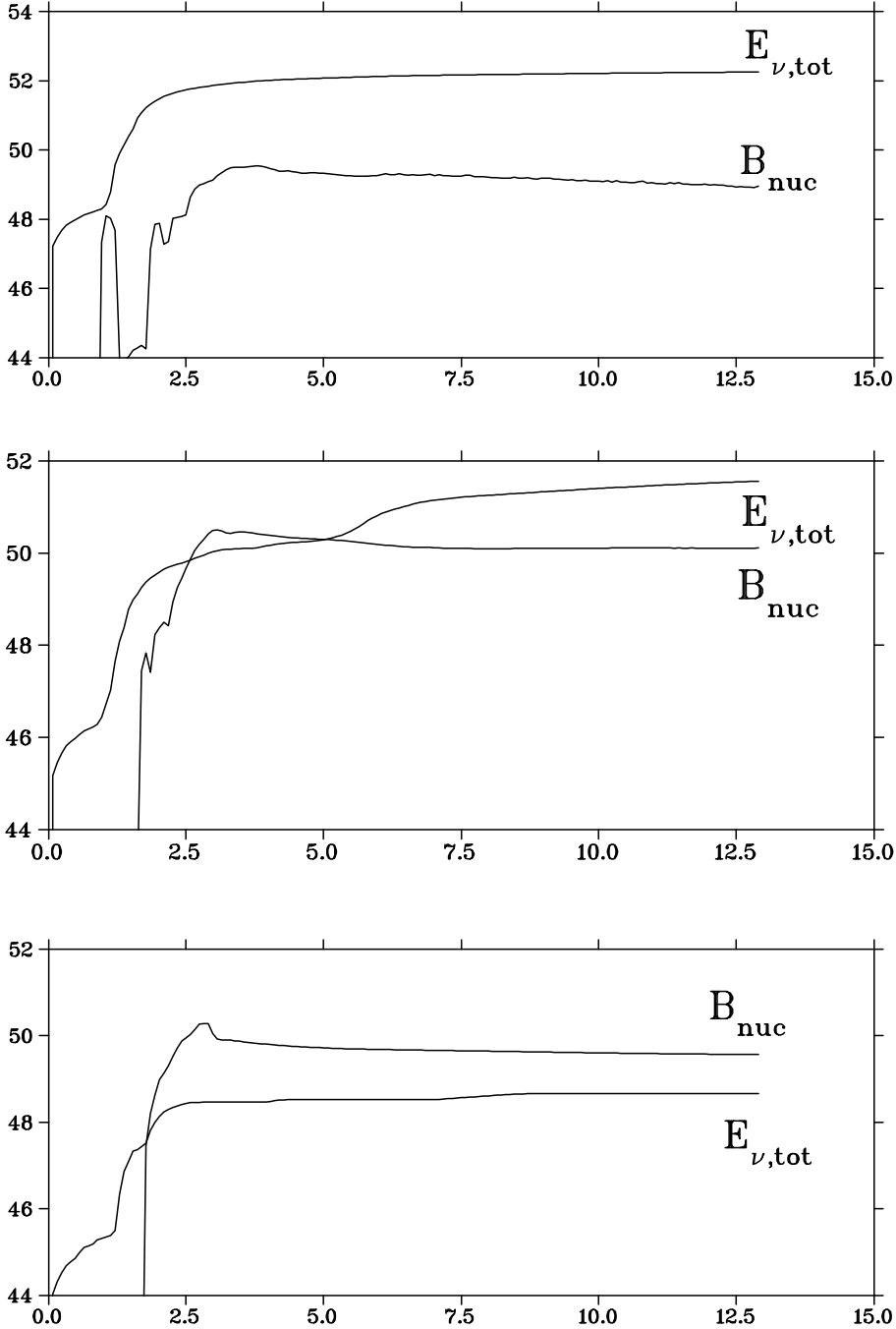,width=12cm,angle=0}
\caption{\label{nunucmult} Comparison of nuclear binding energy and the total energy emitted in neutrinos for the three morphological regions of the coalesced object. The upper panel refers to the central object, the one in the middle to the disk and the lowest to the spiral arms. On the abscissa time (in ms) and on the ordinate the logarithm of the energies (in erg; B$_{{\rm nuc}}$ and E$_{\nu,{\rm tot}}$ denote nuclear binding energy and the total emitted neutrino energy) is shown. In the central object (in this simple model) the neutrino emission dominates clearly over the released nuclear binding energy. The "bump" in panel one short before contact results from nuclei that form when the density decreases due to tidal stretching. In the low mass spiral arms (see Table \ref{masses}) clearly the nuclear energy deposition ($\sim 8 \cdot 10^{49}$ erg) dominates. }
\end{figure} \clearpage

\begin{figure}[h]
\psfig{file=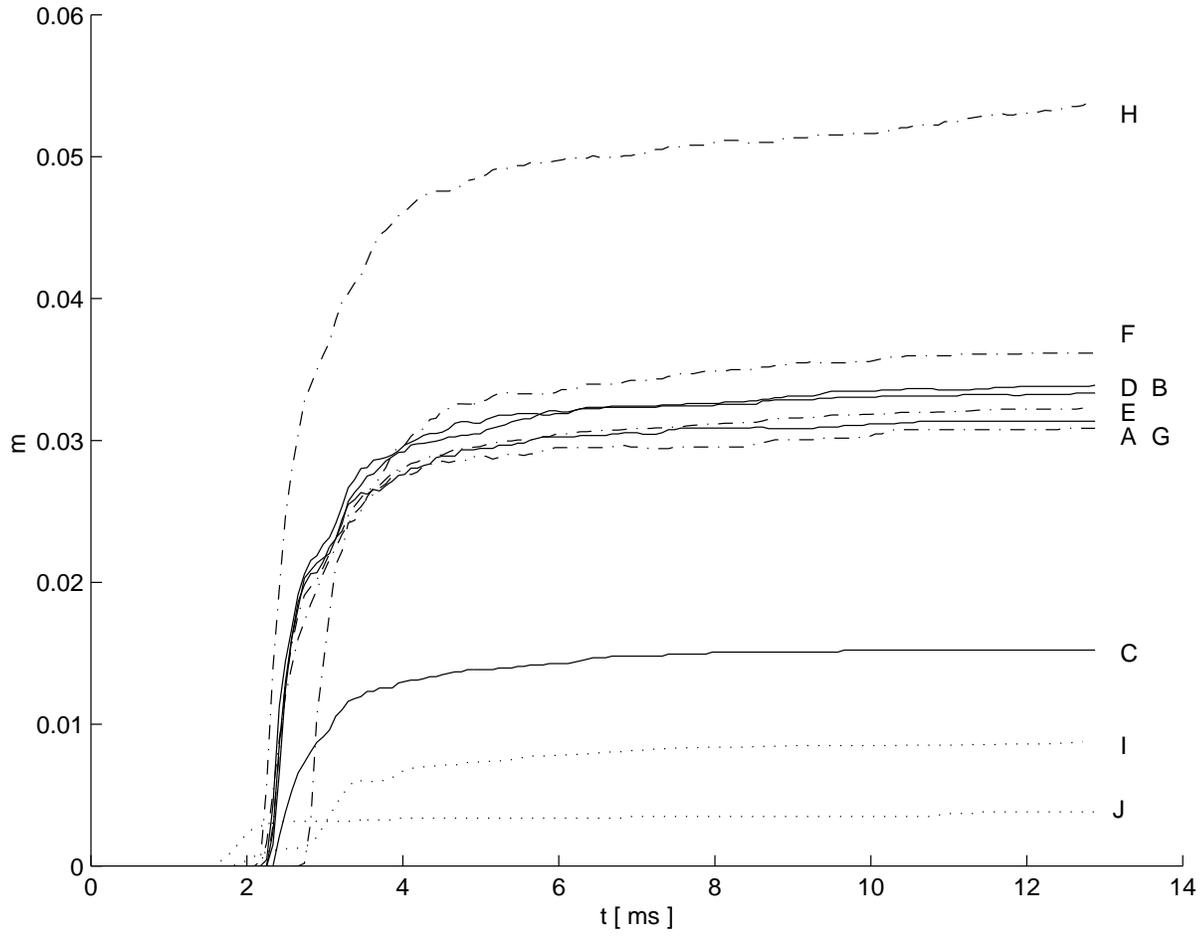,width=16cm,angle=0}
\caption{\label{mesct} Mass ejection (in solar units) of the different models.}
\end{figure} \clearpage

\begin{figure}[h]
\psfig{file=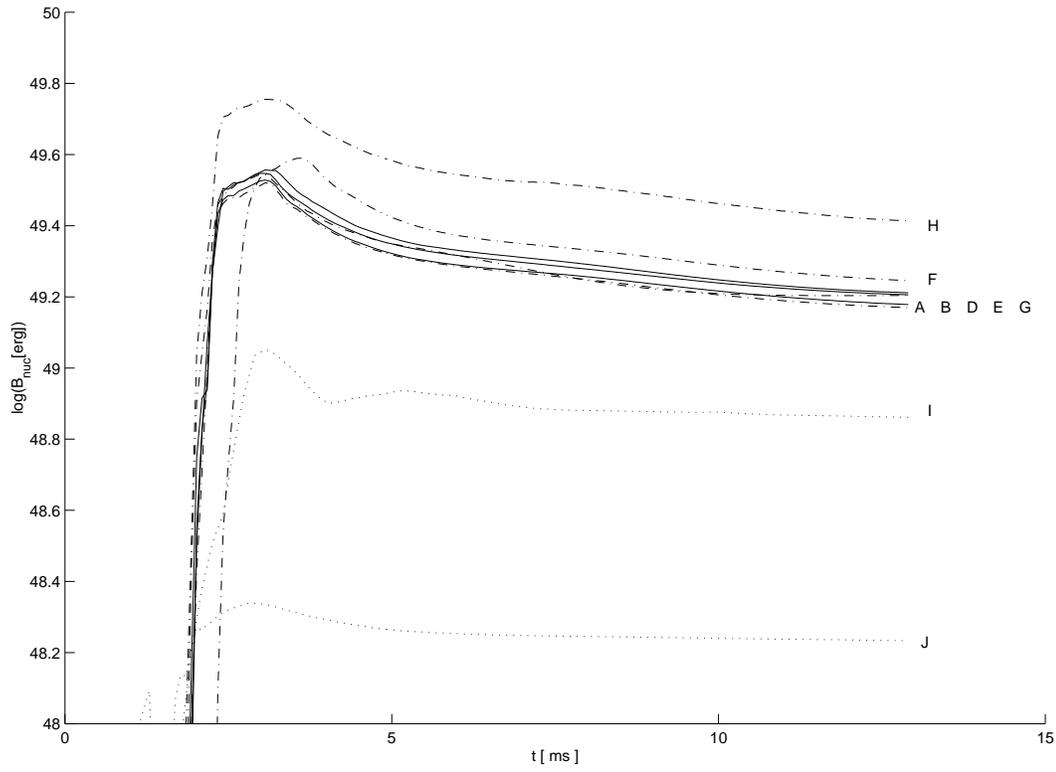,width=14cm,angle=90}
\caption{\label{bnucesc} Logarithm of total nuclear binding energy (in erg) present in the escaping mass. }
\end{figure} \clearpage

\begin{figure}[h]
\psfig{file=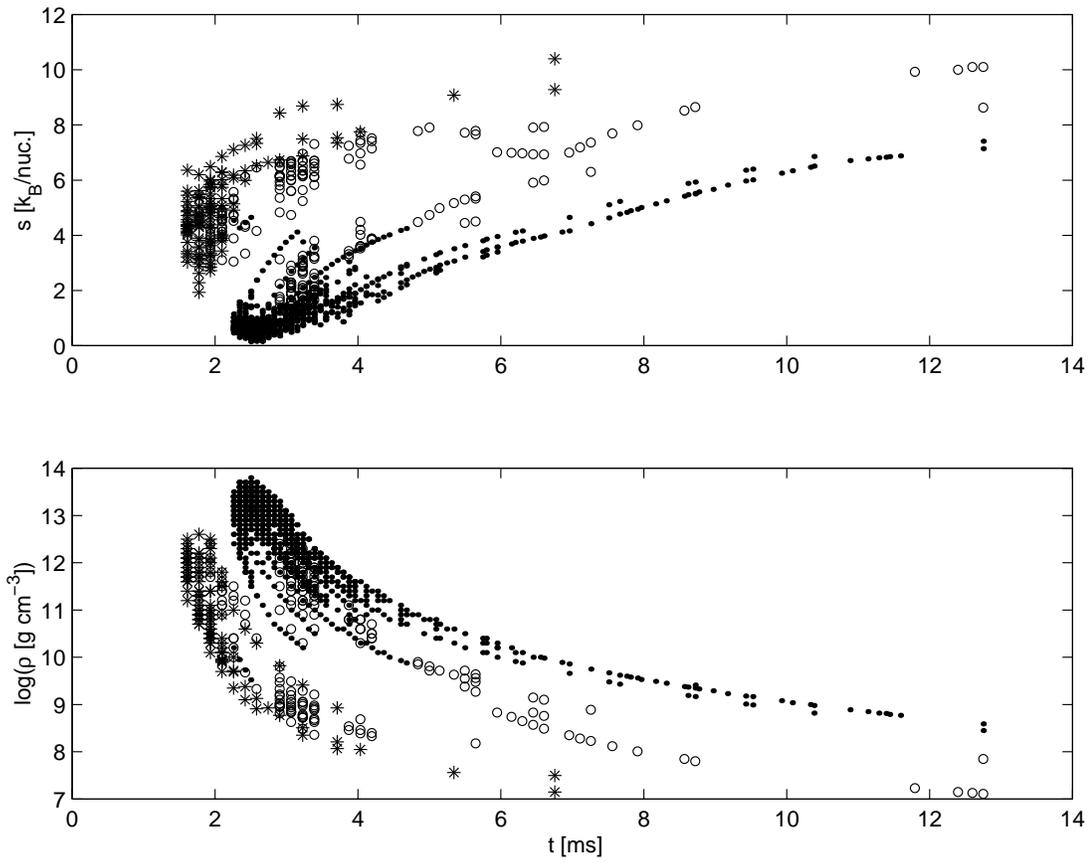,width=16cm,angle=0}
\caption{\label{escprop} Entropies and densities  of the  particles at the moment of ejection. Dots refer to  run E (corotation), open circles to run I (no spin) and asterisks to run J (spins against orbit). }
\end{figure} \clearpage

\begin{figure}[h]
\psfig{file=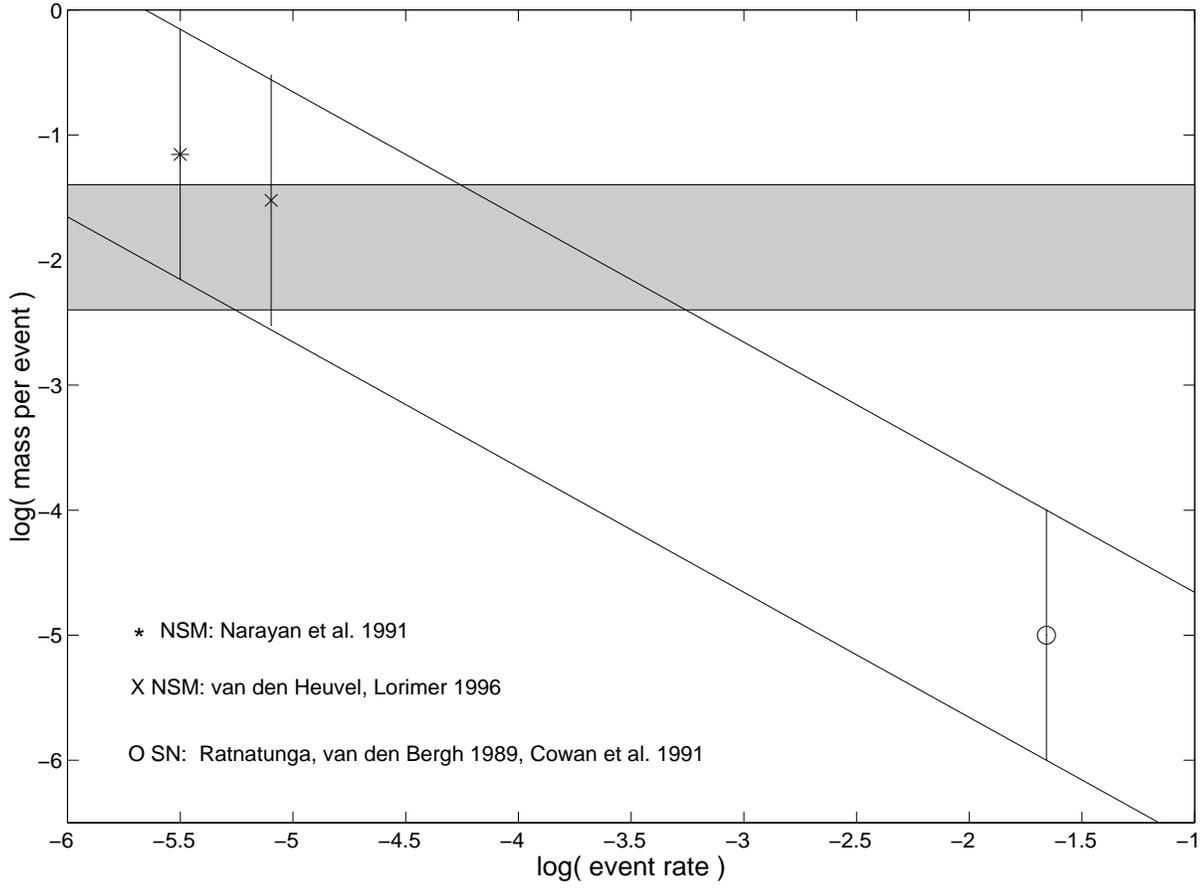,width=16cm,angle=90}
\caption{\label{ratecomp} The shaded region shows the amount of ejected material found in our calculations. The circle shows the amount of ejecta per event if SN II are assumed to be the only sources of the r-process. The asterisk gives the needed ejecta per merging event for the rate of Narayan et al. (1991), the cross for the estimate of van den Heuvel and Lorimer (1996). The event rate is given in year$^{-1}$ galaxy$^{-1}$, the ejected mass in solar units.}
\end{figure} \clearpage

\begin{figure}[h]
\psfig{file=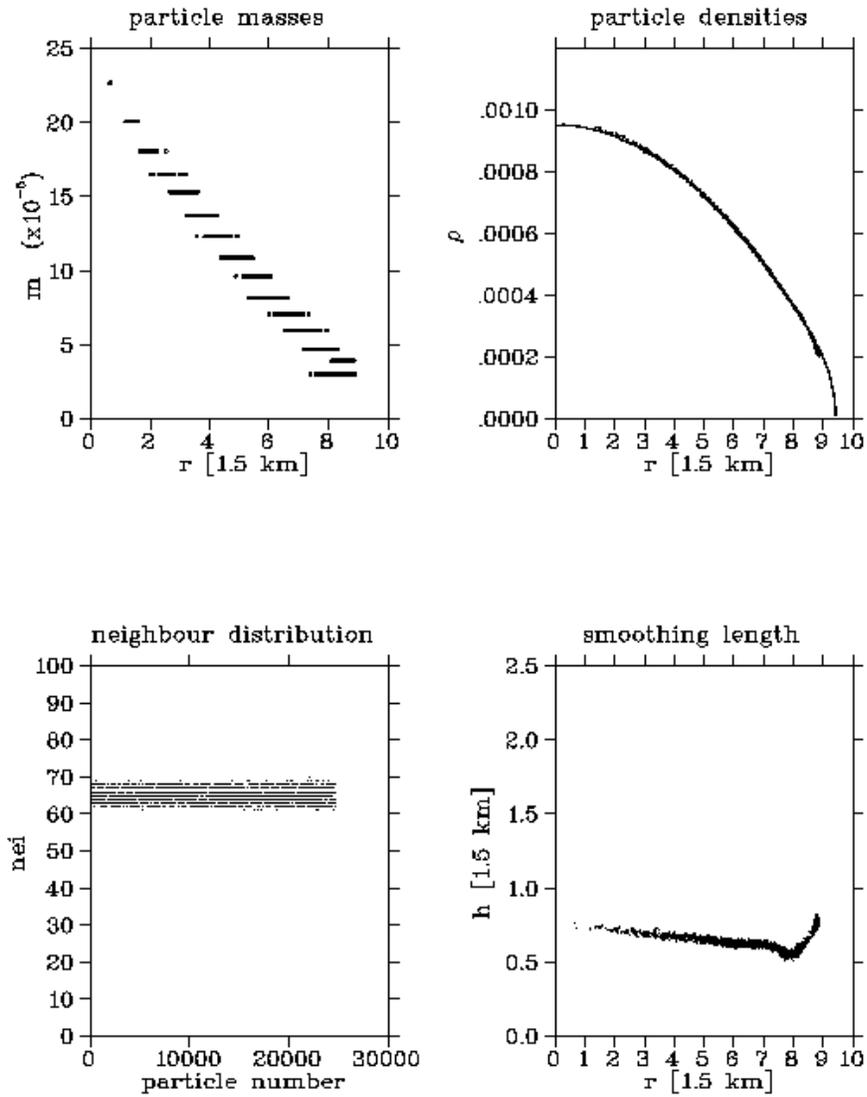,width=14cm,angle=0}
\caption{\label{finsetup} Properties of the configuration after relaxation. The upper left (upper right, lower right) panel shows the distribution of the particle masses (densities, smoothing lengths) with radius, the lower left panel shows the neighbour numbers of each particle.}
\end{figure} \clearpage
\end{center}

\end{document}